\documentclass[
journal=apchd5, 
manuscript=perspective, 
layout=traditional,
]
{achemso}

\usepackage{amsmath,amssymb}
\usepackage{graphicx}
\usepackage{float}
\usepackage{natbib}
\usepackage{color}
\usepackage{xcolor}
\usepackage[colorlinks=true,bookmarks=false,citecolor=blue]{hyperref}

\graphicspath{ {./Figures/} }

\newcommand{\br}{\mathbf{r} }
\newcommand{\bk}{\mathbf{k}  }
\newcommand{\bE}{\mathbf{E} }
\newcommand{\bD}{\mathbf{D} }
\newcommand{\bH}{\mathbf{H} }
\newcommand{\bB}{\mathbf{B} }
\newcommand{\bG}{\mathbf{G} }

\newcommand{\bp}{\mathbf{p} }

\newcommand{\bs}{\mathbf{s} }

\newcommand{\ux}{\widehat{\mathbf{x}} }
\newcommand{\uy}{\widehat{\mathbf{y}} }
\newcommand{\uz}{\widehat{\mathbf{z}} }

\newcommand{\eps}{\varepsilon }
\renewcommand{\Re}{\frak{R}\mathrm{e} }
\renewcommand{\Im}{\frak{I}\mathrm{m} }

\newcommand{\dyad}[1] {\overset\leftrightarrow{\mathbf{#1}}}

\newcommand{\elm}{electromagnetic }
\newcommand{\fp}{Fabry-P\'erot }

\newcommand{\rot}{\nabla \times }

\title{Towards chiral polaritons}

\author{Denis G. Baranov}
\affiliation{Center for Photonics and 2D Materials, Moscow Institute of Physics and Technology, Dolgoprudny 141700, Russia}
\email{denis.baranov@phystech.edu}

\author{Christian Sch\"afer}
\affiliation{MC2 Department, Chalmers University of Technology, Sweden}
\email{christian.schaefer.physics@gmail.com} % since the perspective part is partially my work it would be nice to have all emails then

\author{Maxim V. Gorkunov}
\affiliation{Shubnikov Institute of Crystallography, FSRC ``Crystallography and Photonics'' Russian Academy of Sciences, Moscow 119333, Russia}
\email{gorkunov@crys.ras.ru}


\begin{document}

\begin{abstract}
Coupling between light and material excitations underlies a wide range of optical phenomena. Polaritons are eigenstates of a coupled system with hybridized wave function. 
Owing to their hybrid composition, polaritons exhibit at the same time properties
typical for photonic and electronic excitations, thus offering new ways for controlling electronic transport and even chemical kinetics. While most theoretical and experimental efforts have been focused on polaritons with electric-dipole coupling between light and matter, in chiral quantum emitters, electronic transitions are characterized by simultaneously nonzero electric and magnetic dipole moments. Geometrical chirality affects the optical properties of materials in a profound way and enables phenomena that underlie our ability to discriminate enantiomers of chiral molecules. Thus, it is natural to wonder what kinds of novel effects chirality may enable in the realm of strong light-matter coupling. Right now, this field located at the intersection of nanophotonics, quantum optics, and chemistry is in its infancy. In this Perspective, we offer our view towards chiral polaritons. We review basic physical concepts underlying chirality of matter and electromagnetic field, discuss the main theoretical and experimental challenges that need to be solved, and consider novel effects that could be enabled by strong coupling between chiral light and matter.
\end{abstract}

\maketitle

\section{Introduction}

A system of a few interacting oscillators, either of classical or quantum nature, is a primitive mathematical model that describes the behavior of a wide spectrum of mechanical, thermodynamic, \elm and other kinds of physical systems.
Polaritons are eigenstates of the system when \elm fields and matter couple strongly, i.e., when the coupling rate exceeds all the decay and decoherence rates in the system, and share a mixed light-matter characteristic.\cite{Mills1974}
%They are characterized by a hybrid wave function with simultaneous presence of the photonic and the matter component \cite{Mills1974}.
In the optical domain, polaritonic states are often realized by means of coupling an optical cavity mode with excitonic or vibrational transitions in resonant media \cite{khitrova2006vacuum, torma2014strong, Baranov2018}.
Owing to this hybrid composition, polaritons show properties indicative of massless photonic and massive electronic/vibrational excitations at the same time \cite{sanvitto2016road}, thus offering new ways for controlling electronic transport and modifying chemical kinetics \cite{hutchison2012modifying,thomas2016ground,galego2016suppressing,Munkhbat2018,feist2018polaritonic,Thomas2019tilting,chen2022cavity}.
%In particular, the possibility of strong light-matter coupling to affect the rates of chemical reactions (ones involving the coupled material as a product or as a reactant) has been the subject of intense experimental \cite{hutchison2012modifying, thomas2016ground, Thomas2019tilting, Munkhbat2018, stranius2018selective, peters2019effect} and theoretical \cite{galego2016suppressing,Herrera2016,feist2018polaritonic,fregoni2022theoretical,schafer2019modification} investigation in the past decade.

Most theoretical and experimental efforts in this field have been focused on coupling optical cavities with linearly- or circularly-polarized electronic transitions of various quantum emitters (QEs). 
This is justified by the abundance of electronic and vibrational emitters whose interaction with light is dominated by the electric-dipole term of the Hamiltonian, while higher-order contributions including the magnetic-dipole term are often minute.  To put it another way, natural materials very weakly respond to magnetic fields oscillating with optical frequencies \cite{Landau1984,merlin2009metamaterials}.
%To put it another way, magnetism of natural materials is weak in the visible range \cite{merlin2009metamaterials}. \Cadd{[Are we sure about this statement? What about ferromagnets, Landau-levels, and all other kinds of materials that can be transiently driven in or out of magnetic states. I think we should be more careful here and either be more specific or rephrase the paragraph.]}
Nevertheless, in certain specific cases, a non-negligible magnetic transition dipole moment can be of pivotal importance. \emph{Chiral} media \cite{Lindell2018,Barron2004} represent a practically relevant example.

%Nevertheless, there are examples of media that exhibit resonances with non-negligible magnetic transition dipole moment. One such practically relevant example is presented by the class of \emph{chiral} media \cite{Lindell2018,Barron2004}. 

A three-dimensional rigid body, which cannot be aligned with its mirror image by a series of rotations and translations, is referred to as chiral \cite{Kelvin1894}.
%The chirality occurs at various scales ranging from the shapes of galaxies down to drug and bio-molecules. 
Two mirrored versions of a chiral object are called left and right enantiomers, and usually qualitative differences between them are obvious. However, an attempt to define a unique quantitative measure of the geometric chirality easily leads to confusion: one can introduce different seemingly meaningful parameters which may vanish for different configurations and remain unrelated to the chiral physical phenomena \cite{Harris1999}.
Much more fruitful is the practical approach, when one assesses the chirality of an object by its interactions with other chiral entities. 

For example, acting on a chiral biological receptor, opposite enantiomers of the same molecule cause different response, perceived as different odor or taste \cite{leitereg1971chemical}. 
In pharmaceutics, the opposite enantiomer of a drug molecule can be useless at best, but often it is insidiously toxic for the chiral human body \cite{mayer1997pharmacodynamics,chhabra2013review}. 
The majority of molecules involved in biological processes, such as neurotransmitters (norepinephrine, ephedrine, and others) are chiral.
In this regard, there is a great demand from the pharmaceutical industry to develop effective methods of separating chiral enantiomers.

Chirality profoundly affects interactions of matter with \elm radiation. As soon as a medium is composed of chiral constituents (such as molecules or meta-atoms), it acquires a magneto-electric response, which mixes \textbf{E} and \textbf{B}, and \textbf{H} and \textbf{D}, correspondingly \cite{Lindell2018}. 
The study of interaction between chiral light and matter has uncovered many fundamental effects, but, so far, has been limited to  weak light-matter coupling, where the presence of the photonic field can be treated perturbatively.
Some recent studies make an attempt to address the behavior of strongly coupled system where at least one of the constituents (either light or the material) is chiral \cite{Baranov2020CD,guo2021optical,zhu2021strong}.
%In the context of polaritonics, it is important that both strongly coupled constituents can be made substantially chiral.
%Thus, transitions of chiral QEs are characterized by a pair of collinear electric and magnetic dipole moments \cite{Barron2004,Tang2010,Klimov2012}. On the other hand, recent progress in strongly chiral metasurfaces enables creating chiral optical cavities \cite{plum2015chiral, Voronin2022}.
All this raises a question: which new phenomena should arise from the chirality of a QE, or of a cavity, or of both, when they become strongly coupled? 
%\Cadd{[The previous 3 sentences don't fit well together. I see what you want to say but maybe try to reformulate them.]}

In this Perspective, we offer a conceptual view on the notion of chiral polaritonics. We review the essential physics underlying strong light-matter coupling, the chirality of matter and electromagnetic fields. We discuss the main theoretical and experimental challenges for chiral polaritonics and provide a perspective on its potential and future development.
%speculate on some of the potential novel effects that can be enabled by strong coupling between chiral light and matter. 
The interdisciplinary nature of the problem suggests that a joint effort of researches from classical electromagnetism, quantum optics, material science, and biology might be needed in order to establish chiral polaritonics as the versatile framework it promises to be.
%succeed in the hunt for chiral polaritons.

%\section{The problem of naive approach}
%\rev{I don't like this. I feel like before we discuss the naive approach, we need to properly explain what is a polariton, handedness, chirality of light, and so on. That \emph{was} a good suggestion, but for now I'd keep this section under the rug.}
%
%Given the present understanding of non-chiral polaritons, it would seem natural that loading an ordinary \fp cavity with chiral matter and following excitation of such a system with circularly polarized \elm field would result in the formation of chiral polaritons. In reality, the situation is a bit less intuitive.  The quasi-normal modes of frequently used cavities - such as \fp resonators - often lack any handedness characteristics of their \elm field. As a result of this achiral geometry of optical cavities, left-handed and right-handed enantiomers of a given material interact with the optical field equally strong, removing any potential asymmetry effects. This difficulty motivates the quest for resonant optical structures supporting QNMs with very particular field polarization characteristics. To see how chirality of light comes into play in the realm of polaritonics, we first revise the basics of strong light-matter interaction, chirality of light and matter, and then show how the two can be naturally joined/combined/brought together such that chirality of light is not lost on the way.

\section{Strong light-matter coupling}

Let us begin by briefly recalling theoretical apparatus 
%basic mathematical frameworks 
describing the interaction of an optical emitter, such as an atom, a molecule, a lattice vibration, or an artificial harmonic meta-atom, with a single \elm mode of an optical cavity.
There are several frameworks to model such an interaction, and the choice of the particular model in each situation depends on the context.
As long as anharmonic effects related to the saturability of two-level system(s) may be ignored, and as long as the coupling strength is not too high compared to resonant energies of the photonic and material oscillators, the light-matter coupling can be conceptually modeled using the following coupled-mode Hamiltonian \cite{khitrova2006vacuum,torma2014strong,Baranov2018}:
\begin{equation}
    H = \left(
    \begin{array}{cc}
    \omega_{cav} - i \gamma_{cav} & g  \\ 
    g^*  & \omega_0 - i \gamma_0
    \end{array} \right),
    \label{Eq_1}
\end{equation} 
where $\omega_0 - i \gamma_0$ and $\omega_{cav} - i \gamma_{cav}$ are the complex-valued frequencies characterizing  the material and photonic eigenmodes, correspondingly \cite{Lalanne2018}, and $g$ is the coupling constant. 
For a single emitter coupled to a cavity we have $g= \mathcal{E}_{vac} \mu$, where $\mathcal{E}_{vac}$ is the vacuum electric field of the cavity mode, and $\mu$ is the transition dipole moment of the emitter. If $N$ identical emitters are coupled to the same cavity mode, the collective coupling constant reads $g = \sqrt{N}  \mathcal{E}_{vac} \mu$ \cite{Baranov2018}.
The two eigenvalues of the Hamiltonian Eq. \ref{Eq_1} read:
\begin{equation}
    \omega_{ \pm} = \frac{\omega_{cav} + \omega_0}{2} - i \frac{\gamma_{cav} + \gamma_0}{2} \pm \sqrt{|g|^2 + \frac{1}{4}(\omega_{cav} - \omega_0 - i (\gamma_{cav} - \gamma_0))^2}.
    \label{Eq_2}
\end{equation}Within this model of light-matter interaction, two different regimes may be realized.

\begin{figure*} [hbt!]
\includegraphics[width=1\textwidth]{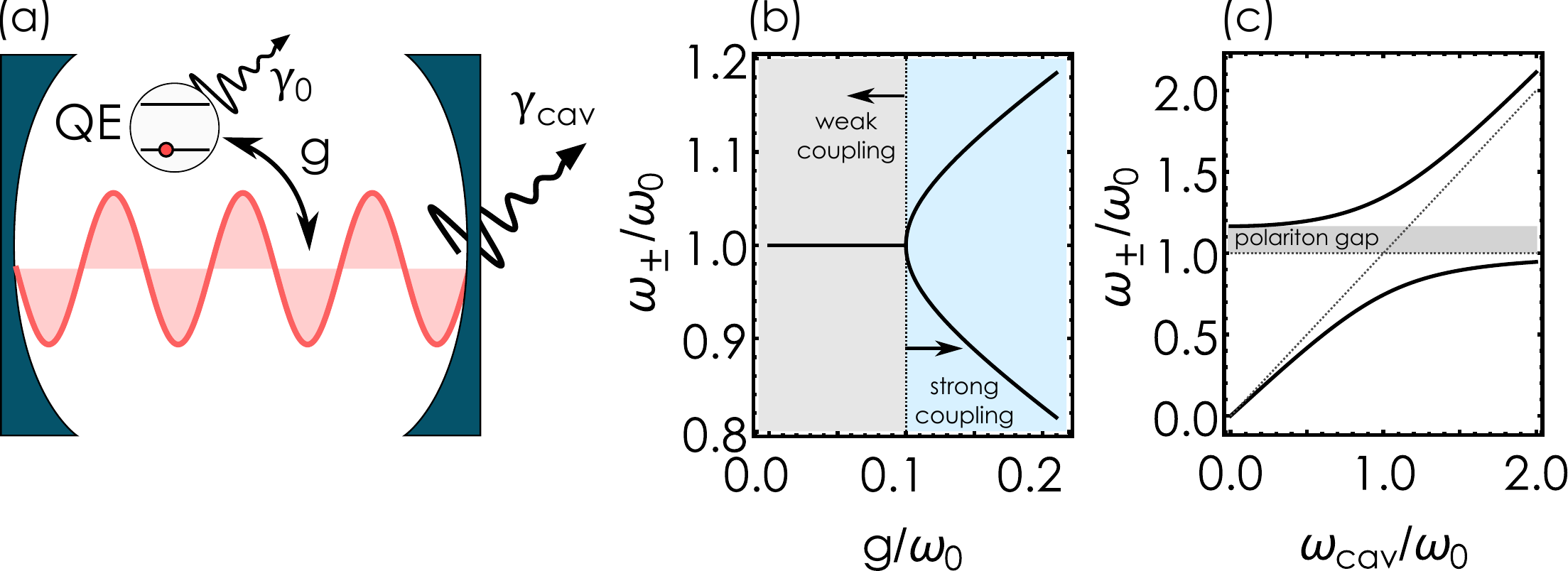}
\caption{\textbf{The basic picture of strong light-matter coupling.} (a) Schematic illustration of an electronic transition of a two-level atom coupled to a single photonic mode of an optical cavity. (b) Eigenvalues of the coupled-mode Hamiltonian, Eq. \ref{Eq_2}, as a function of the coupling constant $g$ for the zero-detuning case, $\omega_{cav} = \omega_0$. (c) The eigenvalues spectrum plotted as a function of the cavity frequency for a fixed coupling constant.}
\label{fig1}
\end{figure*}

Fig. \ref{fig1}(b) shows the eigenvalues of Hamiltonian Eq. \ref{Eq_1} as a function of the coupling constant $g$ for the zero-detuning case, $\omega_{cav} = \omega_0$.
Weak coupling regime occupies the left part of this diagram, where $g < |\gamma_{cav} - \gamma_0|/2$.
In this regime the two eigenstates have equal real parts of their eigenenergies (although different imaginary parts). An excited QE undergoes exponential decay into the ground state at a rate $\gamma$, accompanied by spontaneous emission of a photon in the environment. 
%The rate of this decay is modified by the presence of the cavity according to $\gamma = F_P \gamma_0$, where $F_P$ is the Purcell factor, which can be calculated given the knowledge of the Green's tensor of the system.
The point $g = |\gamma_{cav} - \gamma_0|/2$ marks the transition to the strong coupling regime.
Once the coupling constant exceeds the threshold value, the
%the eigenvalues of the coupled system experiences a qualitative change. 
real parts of eigenenergies split into two branches separated by the so called Rabi splitting. Correspondingly, wave functions of the two eigenstates become equal 50\% mixes of photonic and material components, and are referred to as upper and lower polaritons.
When plotted as a function of the cavity frequency for a fixed coupling constant, the energy spectrum displays the well-known picture of anti-crossing between two polaritonic modes, Figure \ref{fig1}(c).% Besides, it features so called polariton gap - a region of energies with no allowed eigenstates within it \cite{Todorov2012}. 

%\Cadd{I would remove all the USC and Hopfield here. The best might be to reduce this section to its bare bone, we only have to transfer the basic idea  and refer to the later section and the Chiral Hopfield paper. DGB - tend to agree, but maybe i'd keep a very short reference to other theoretical approaches}

When the coupling constant becomes comparable to the resonant energies of the system  ($g>0.1\omega$), the system enters the ultrastrong coupling (USC) regime in which all eigenstates become noticeably renormalized by the interaction\cite{FriskKockum2019}. At this point, the trivial coupled-mode Hamiltonian as well as the widely used rotating-wave approximation yield unphysical eigenstates and becomes inaccurate, thus the underlying theoretical foundation of QED has to be consulted. A simple and widely used model was derived by Hopfield \cite{hopfield1958theory, woolley2020power}.
In coupled structures where the spatial coherence of the polariton is at the center -- for example, in systems exhibiting polariton condensation \cite{daskalakis2014nonlinear,sanvitto2016road} -- other approaches might be used that capture the non-local and non-linear properties of polaritons, such as the Gross-Pitaevskii equation.

%Besides systems with physically separated cavity mirrors and resonant material, polaritonic states can emerge in monolithic structures, where the photonic mode and the electronic or vibrational transition originate from the same material. This is the case of so called cavity-free polaritons \cite{canales2021abundance,thomas2021cavity}, where the photonic mode emerges thanks to partial light reflections at the boundaries of the object with background permittivity $\eps_{\infty} > 1$.

The hybrid wave function of polaritons has far-reaching implications for the microscopic properties of the coupled system.
In particular, the possibility of strong light-matter coupling to affect the rates of chemical reactions (ones involving the coupled material as a product or as a reactant) has been the subject of intense experimental \cite{hutchison2012modifying, thomas2016ground, Thomas2019tilting, Munkhbat2018, Stranius2018, peters2019effect} and theoretical \cite{galego2016suppressing,Herrera2016,feist2018polaritonic,schafer2019modification,schafer2021shining,fregoni2022theoretical,PhysRevLett.128.096001} investigation in the past decade.
Thus, it is natural to wonder whether bridging polaritons with chirality may break the symmetric yield of chiral products and enable other practically valuable effects.

\section{Field and matter chirality}

\subsection{Chirality of light}

\begin{figure*}[hbt!]
\includegraphics[width=.9\textwidth]{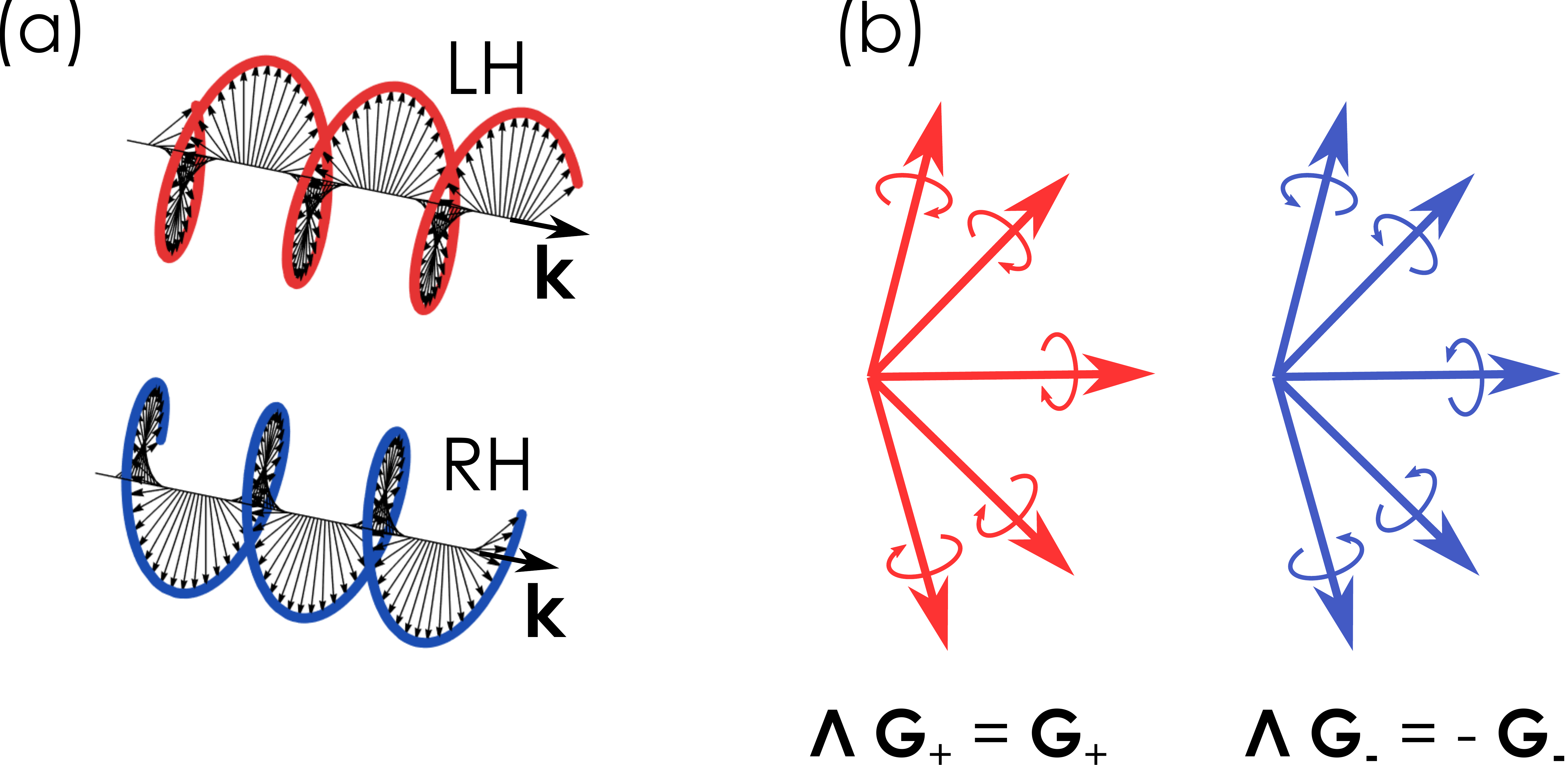}
\caption{\textbf{The basics of \elm field chirality.} (a) Illustration of the spatial structure of and RCP and LCP plane wave in free space. The arrows indicate spatial distribution of the instantaneous electric field vectors in space. (b) Schematic illustration of expansion of an arbitrary non-paraxial \elm field into Riemann-Silberstein vectors.}
\label{fig3}
\end{figure*}

In the following subsections we briefly overview the concepts of chiral light and chiral matter. Chirality of light is most easily approached through the notion of \emph{handedness}.
%In the context of chiral \elm field the notion of \emph{handedness} will be useful.
Left-handed (LCP) and a right-handed (RCP) circularly polarized plane waves represent primordial examples of chiral \elm field, Fig. \ref{fig3}(a).
 LCP (RCP) wave in a transparent dielectric can be defined as a monochromatic plane wave with its magnetic field $\pi/2$ ahead (behind) the electric field everywhere in space:
\begin{equation}
    Z\bH _{LCP}(\br) = -i \bE_{LCP}(\br),\ \
    Z\bH _{RCP}(\br) = +i \bE_{RCP}(\br).
    \label{Eq_6}
\end{equation} 
where $Z = \sqrt{\mu\mu_0/ \eps\eps_0}$ is the medium impedance. Considering an RCP (LCP) wave frozen in time yields a helix twisted clockwise (counter-clockwise) in the propagation direction, Fig. \ref{fig3}(a).
The handedness quality of a circularly polarized plane wave is preserved upon arbitrary rotation of the field in space, and also upon time reversal. However, this quality is flipped upon a mirror reflection, or an inversion. Thus, the handedness can be used as a basis for constructing a time-even pseudo-scalar field. 
%One can wonder: Is it possible to introduce a quantitative measure of handedness applicable to  arbitrary states of free-space \elm field?

%\rev{[For circular polarizations only? I am afraid, handedness is not a quantity but rather a qualitative term. Helicity is the right quantity to characterize the light chirality.]} 

To that end, consider the helicity operator $\dyad{\Lambda} = \frac{1}{k}\big(\begin{smallmatrix}
  \rot & 0\\
  0 & \rot
\end{smallmatrix}\big)$
acting on the complex electric and magnetic field vectors $\Psi = (\bE, Z\bH)^T$.
One immediately realizes that LCP and RCP waves are eigenstates of this operator with eigenvalues $\lambda = + 1$ and $\lambda = - 1$, respectively \cite{corbaton2014helicity, Fernandez-Corbaton2016}. Correspondingly, any linear combination of LCP (RCP) plane waves is an eigenvector of $\dyad{\Lambda}$.
%This observation suggests a way towards a local measure of handedness of a non-homogeneous field. 
Calculating the local expected value of the helicity operator $\langle \Psi(\br) \rvert \dyad{\Lambda} \rvert \Psi(\br) \rangle$ and adjusting the prefactor, we obtain the chirality density -- a local measure of handedness of a non-homogeneous field -- that was originally introduced by Lipkin \cite{Lipkin1964}:
\begin{equation}
    C(\br,\omega) = \frac{\eps_0\omega }{2} \Im[\bE\cdot \bB^*].
    \label{Eq_8}
\end{equation}
This quantity characterizes the local geometry of light polarization, expressing the degree of ``screwiness''of the electric and magnetic field lines \cite{Yang2011}. It can also be interpreted as the normalized difference between the number of left-handed and right-handed photons in a monochromatic field.
%similarly to the spin angular momentum, which measures the degree of rotation of electric and magnetic fields. However, the two quantities are strikingly different, and we will discuss this difference in details later in this section.
On a side note, the presence of material dissipation strongly modifies the expression for chirality density \cite{Vazquez-Lozano2018}.

To grasp how the chirality density is generally applicable to arbitrary monochromatic \elm field distributions, one notices that a solution of Maxwell's equations in a transparent medium can always be expanded into two components with well-defined handedness, called Riemann-Silberstein vectors \cite{bialynicki2013role}:
\begin{equation}
    \bG_{\pm }(\br) = \frac{\bE \left( \br \right) \pm iZ \bH  (  \br) }{\sqrt{2}}.
    \label{Eq_9}
\end{equation} 
Each of the two Riemann-Silberstein vectors is an eigenstate of the helicity operator, meaning that it only contains fields of the corresponding handedness. 
According to our handedness convention, $\mathbf{G}_+$ ($\mathbf{G}_-$) represents the LH (RH) field component. 

%\rev{In spite of its simplicity and seeming universality, the density \eqref{Eq_8} is a local quantity, i.e., by its definition, insensitive  to the spatial field distribution, and thus incapable of exhaustive characterization of all possible chiral light shapes. Indeed, defining an \elm field to be chiral in the most general way if its mirror image cannot be aligned with itself by a series of rotations and translations, one quickly realizes that there are plenty of chiral solutions of Maxwell's equations with identically zero $C(\br,\omega)$. For example, numerous chiral vortex beams that can be build of \elm fields with local linear polarization combine chirality with  $C=0$.\cite{nechayev2021kelvin} Although it is unclear at the moment whether the distribution chirality, also dubbed as Kelvin's chirality, can be quantified by an appropriate measure, various possibilities of its conversion from and into the local chirality density in tightly focuses of light beams are being intensively explored.\cite{forbes2021orbital,forbes_optical_2022} }

It should be noted that, in the most general sense, a combination of \elm fields is chiral, if its mirror image cannot be aligned with itself by a series of rotations and translations. 
Clearly, the local density \eqref{Eq_8}, being perfect for characterizing plane waves, cannot exhaustively cover the structured light chirality. 
There are plenty of solutions of Maxwell's equations, such as chiral vortex beams, having $C = 0$ everywhere in space, yet being geometrically chiral.\cite{nechayev2021kelvin}. 
It is unclear at the moment whether this distribution chirality, also dubbed as Kelvin's chirality, can be quantified by an appropriate measure, but various ways of its conversion from and into the local chirality density in tight focuses of light beams are being intensively explored.\cite{forbes2021orbital,forbes_optical_2022}. When it comes to chiral light-matter interactions, however, it is the chirality  density \eqref{Eq_8} that still plays the major role. Peculiar effects determined by the distribution chirality, though being more subtle, promise intriguing new possibilities along with the surge of interest in structured light in general.

%For example, for an RCP plane wave $\mathbf{G}_+ = 0$ at every point in space.
%\rev{[The following could be safely trimmed if needed.]}
%Furthermore, it can be shown that the position-dependent difference between their norms is proportional to the chirality density \cite{corbaton2014helicity}:
%\begin{equation}
%    C(\br)\propto {\left|\bG_+ ( \br)\right|}^2- {\left|\bG_- ( \br)\right|}^2.
 %   \label{Eq_10}
%\end{equation} 

%Using the set of Riemann-Silberstein vectors the curl Maxwell's equations in free space can be rewritten as \cite{corbaton2014helicity}:
%\begin{align}
 %   \rot{\bG_{+}} = \frac{i}{c} \frac{\partial \bG_{+}}{\partial t},\ \
 %   \rot{\bG_{-}} = \frac{i}{c} \frac{\partial \bG_{-}}{\partial t}.
%\end{align}
%This form of Maxwell's equations highlights the kind of decoupling between vector fields which occurs when the \elm field is treated in the basis of Riemann-Silberstein vectors as opposed to the vectors of electric and magnetic fields.

\subsection{Chirality of matter}

\begin{figure*}[t]
\includegraphics[width=1\textwidth]{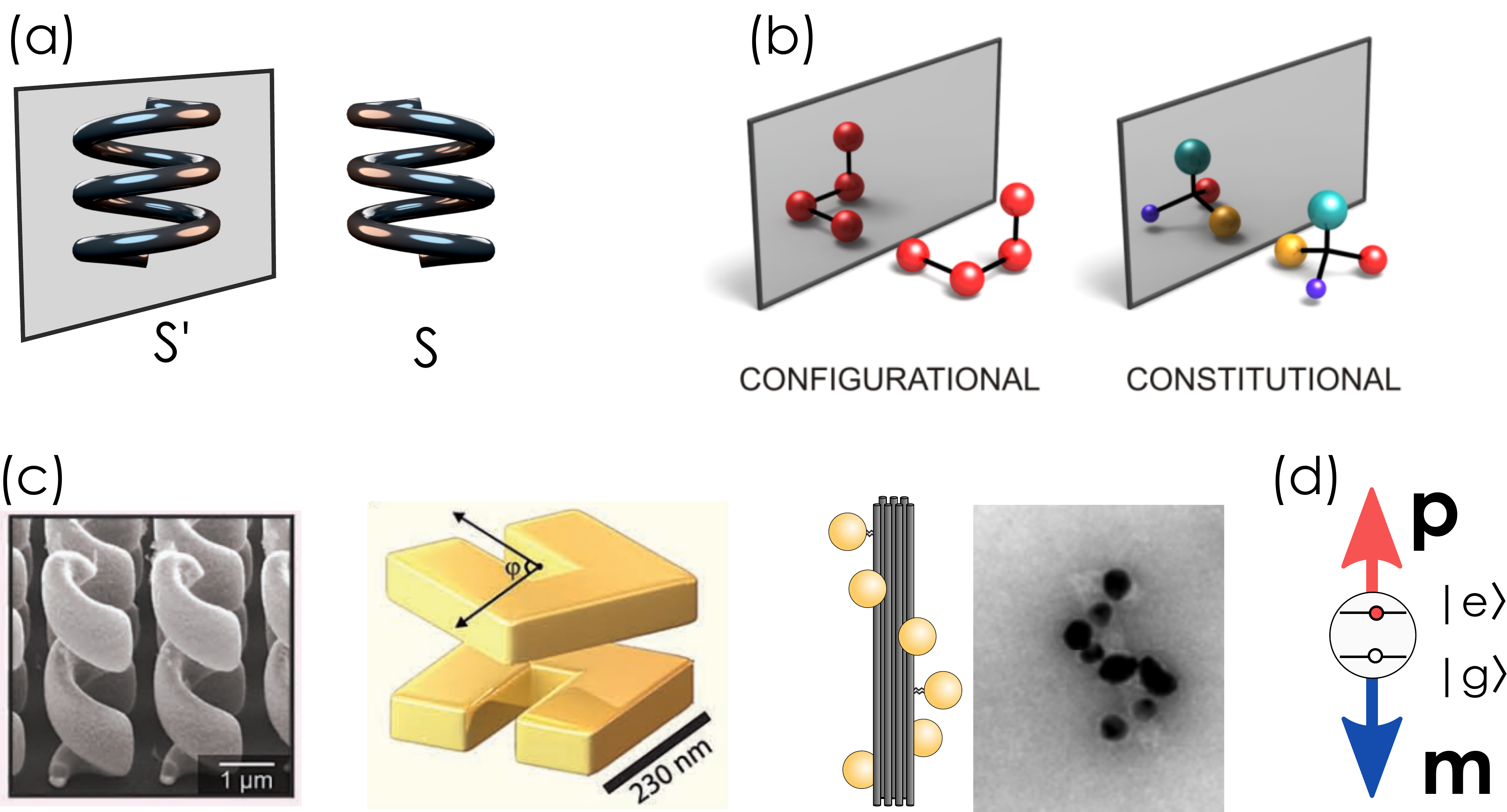}
\caption{\textbf{The basics of matter chirality.} (a) Visual representation of a chiral object $S$ and its mirrored enantiomer $S'$. The two objects are related by a mirror transformation and cannot be mapped into one another by a series of rotation and translations. (b) Configurational and constitutional chirality. Configurational chirality involves identical constituent elements arranged in a geometrically chiral shape. Constitutional chirality involves non-identical elements arranged in an otherwise non-chiral shape. (c)  Examples of chiral meta-atoms: metallic helices, twisted split-ring resonators, and DNA origami scaffolds. (d) Schematic illustration of a chiral source of \elm field involving a pair of parallel electric and magnetic dipole transitions.
Panel (b) adapted with permission from refs. \cite{hentschel2012three}.
Parts of panel (c) adapted with permission from refs. \cite{noguez2009optically} and \cite{kuzyk2012dna}. }
\label{fig2}
\end{figure*}

Not only \elm field, but rigid geometrical bodies as well can be characterized with chirality.
A geometrical body in three-dimensional space is said to be chiral, if it cannot be superimposed upon its mirror image by rotations and translations, Fig. \ref{fig2}(a).
Geometrical chirality has fundamental implications on the \elm response of matter. Any medium composed of chiral microscopic elements, such as molecules or meta-atoms, acquires a magneto-electric response, wherein electric displacement $\bf D$ becomes linked to $\bH$, and magnetic induction $\bf B$ becomes linked to $\bE$.
For a reciprocal bi-isotropic material the constitutive relations take the form \cite{Lindell2018}:
\begin{equation}
    \left( \begin{array}{c}
    \bD  \\ 
    \bB  \end{array}
    \right) = \left( \begin{array}{cc}
    \eps_0  \eps(\omega) & +i\kappa(\omega) /c \\ 
    -i\kappa(\omega) /c & \mu_0 \mu(\omega)  \end{array}
    \right)\left( \begin{array}{c}
    \bE  \\ 
    \bH  \end{array}
    \right),
    \label{Eq_5}
\end{equation} 
where $\eps$ and $\mu$ are the scalar relative permittivity and permeability, correspondingly, and $\kappa=\kappa'+i\kappa'' $ is a complex dimensionless pseudoscalar \emph{Pasteur parameter} of the coupling between the electric and magnetic fields. 
Generally, any of these quantities can be a tensor (pseudotensor), in which case the medium is called bi-anisotropic.
In the case of a bi-isotropic medium with a pseudoscalar Pasteur parameter, its real part $\kappa'$ introduces a difference in phase velocities of RCP and LCP plane waves, resulting in the optical rotation (OR) of linear polarization direction. The imaginary part $\kappa''$ gives rise to different absorption of RCP and LCP waves, which amounts to the circular dichroism (CD). 
%\Cadd{[Maybe we shoud explicitly write the frequeny dependence to emphasize that the Pasteur parameter is also frequency dependent (to contrast Feist's paper). $\eps$ is also tensorial.]}

Prominent examples of naturally chiral compounds are different inorganic crystals such as quartz or cinnabar. %\Cadd{[Which are also anisotrop and therefore $\eps_{ij}$! I think we should be as general as possible in (3).]} Their structural chirality is determined by a helical arrangements of atoms along certain crystallographic directions. 
%Observing the corresponding chiral optical phenomena (OR or CD) in such crystals requires caution and precision. First, one has to exclude much stronger effects of the crystal birefringence by adjusting the light propagation direction along an optical axis. Second, the parameter $\kappa$ is so small that it requires the light to travel many thousands of wavelengths to accumulate measurable difference in phase delay or absorption \cite{Nye1985}.
Much more common is the chirality on the molecular scale of organic substances. Although the corresponding effects of bi-isotropy described by Eq.~\eqref{Eq_5} as well as of a more sophisticated bi-anisotropy remain also typically weak, the valuable information encoded in OR and CD spectra motivates persistent improvement of the corresponding techniques of optical chirality diagnostics \cite{Polavarapu2018}.

Profound chiral optical phenomena occur when the molecular chirality is given an opportunity to govern a bottom-up self-assembly of chiral superstructures. For a particular example, one can consider chiral nematic liquid crystals -- cholesterics. Here the structural chirality of constituent organic molecules induces smooth twisting of the otherwise homogeneous nematic. Upon appropriate preparation and alignment, cholesterics form stable helical configurations with
variable pitch \cite{Blinov2011}.
%in the range from several hundreds of nanometers to tens of micrometers, depending on the chiral dopant concentration. From the optical point of view, such superstructures are helically twisted birefringent transparent media which absolutely differently scatter the RCP and LCP light: While waves of one circular polarization experience the birefringent helix as a uniform medium with an average refractive index, the others undergo coherent reflection from the periodic modulations resulting in formation of chiral photonic stop-bands. Notably, such extremely chiral properties are exhibited by the layers of a thickness of about tens of micrometers, i.e., by several orders of magnitude more compact compared to inorganic chiral crystals 

Besides the variety of naturally occurring organic and inorganic substances, chirality can be engineered in nanostructures at the mesoscopic level \cite{ma2017chiral,Cecconello2017,hentschel2017chiral}.
%artificial/engineered nanostructures offer an endless 
Truly fascinating prospects of boosting the chiral optics to its fundamental limits have opened with the advent of metamaterial concept, as the most sophisticated top-down nanotechnology approaches have been applied to produce designer meta-structures for versatile optical functionality \cite{Schaeferling2017}. 
Typically, their chirality stems from that of constituting elements -- \emph{meta-atoms}. The mirror symmetry of the  latter can be broken in different ways. 
%Chirality of artificial meta-atoms may have various origin. 
A meta-atom chirality may originate purely from the geometry of its construction, in which case, it is referred to as \emph{configurational} chirality, Fig. \ref{fig2}(b). Alternatively, it may be determined by the mutual arrangement of specific different parts of an otherwise achiral structure, in which case it is referred to as \emph{constitutional} chirality.
Examples of artificial chiral nanostructures include metallic helices \cite{schaferling2012tailoring,schaferling2014helical}, chiral arrangements of non-chiral meta-atoms \cite{liu2009stereometamaterials}, and numerous bio-inspired chiral plasmonic nanostructures build around DNA scaffolds \cite{Cecconello2017,avalos2022chiral,Kuzyk2012}, Fig. \ref{fig2}(d).  It is truly remarkable how almost infinitesimally small bio-molecules can impart chiral optical response to otherwise achiral plasmonic structures. The latter can amplify and effectively transfer subtle chiral fingerprints of bio-molecules from the ultraviolet to the more convenient visible range \cite{Lu2013}. Moreover, proteins can trigger the formation of chiral plasmonic complexes which then directly visualize the structural molecular chirality by the visible plasmonic CD \cite{Zhang2019}.

%For a subwavelength chiral reciprocal meta-atom the induced electric $\bf p$ and magnetic $ \bf m $ dipole moments are related to background electric $\bE$ and magnetic $\bH$ fields via \cite{corbaton2014helicity, Fernandez-Corbaton2016}:
%\begin{equation}
%\left( \begin{array}{c}
 %   \bp\\ 
  %  \mathbf{m}  \end{array}
   % \right) = \left( \begin{array}{cc}
    %\eps_0{\alpha }_e & +i{\alpha }_{em}/c \\ 
    %-i\eps_0c{\alpha }_{em} & {\alpha }_m \end{array}
%    \right)\left( \begin{array}{c}
 %   \bE  \\ 
  %  \bH  \end{array}
   % \right),
%\end{equation} 
%where ${\alpha }_e$, ${\alpha }_m$, and ${\alpha }_{em}$ are the electric, magnetic, and magneto-electric dipole polarizabilities, respectively.

Metasurfaces -- periodic arrangements of chiral meta-atoms in planar arrays -- have proven to perform extremely strong chiral light-matter interactions within the optical paths smaller and even much smaller than the wavelength \cite{Kim2021}. Moreover, the palette of available chiral optical phenomena greatly exceeds the capabilities of natural materials: a chiral metasurface is characterized by a set of scattering matrix elements $r_{\mu \nu}$ and $t_{\mu \nu}$ describing transmission and reflection of light from polarization state $\nu$ into polarization state $\mu$.
Depending on the metasurface design, some of these coefficients can vanish, while others become dominant. The metasurface point symmetry group plays the key role in the analysis of all such possibilities.

%Thus, in addition to the co-polarized transmission coefficients $t_{RR}$ and $t_{LL}$ of the RCP and LCP waves, characterizing the Pasteur medium, a metasurface can perform transformations of circular polarizations described by the cross-polarized coefficients $t_{RL}$ and $t_{LR}$. For the reflections, a similar set of co-polarized and cross-polarized reflection coefficients $r_{RR}$, $r_{LL}$, $r_{RL}$, and $r_{LR}$ is introduced.   

Chiral metasurfaces possessing a three-fold or higher rotational symmetry around a vertical axis scatter normally incident light in a relatively simple fashion \cite{Menzel2010, Kondratov2016}.
The only relevant chiral characteristics of such metasurfaces are the transmission CD, which describes the difference in transmittances of RCP and LCP waves:
\begin{equation}\label{CD}
    \textrm{CD} = \frac{|t_{RR}|^2 - |t_{LL}|^2}{|t_{RR}|^2 + |t_{LL}|^2},
\end{equation}
and OR, which describes the difference in phase delays.
Qualitatively, such metasurfaces are similar to slabs of natural chiral Pasteur media \cite{Bai2007}. The quantitative differences, however, are immense: the CD and OR delivered by metasurfaces can take any physically meaningful value including the extreme maxima and minima (${\rm CD}=\pm 1$ and ${\rm OR}=\pm 90^\circ$) after the light travels only  fractions of its wavelength \cite{Gorkunov2014}. Recent transition to all-dielectric platforms has enabled creating strongly chiral and transparent samples \cite{Zhu2018,Gorkunov2018,Tanaka2020}.
%Experimental silicon metasurfaces with a $C_4$ rotational axis have proven to transmit up to 0.9 of waves of one circular polarization with the CD reaching 0.7 \cite{Tanaka2020}. 
Moreover, precisely engineering  quasi-bound states in the continuum \cite{Hsu2016} it is possible to realize the \textit{maximum chirality} combining full transparency to waves of one circular polarization with full absorption of waves of the opposite circular polarization \cite{Gorkunov2020Metasurfaces}. 

Combining chirality with broken rotational symmetry allows obtaining even more sophisticated polarization responses, which expands the multitude of chiral metasurface functionalities further \cite{wu2014spectrally,wang2016circular,Wang2021,gautier2022planar}.
%Less symmetric chiral metasurfaces can perform even more unnaturally, as their CD is not necessarily determined by the chiral absorption. 
Carefully engineered arrays of silicon helices \cite{Karakasoglu2018} or silicon bars of different height \cite{gorkunov2021bound,kuhner_unlocking_2022} achieve lossless maximum chirality, when transparency to waves of one circular polarization is accompanied by a full reflection of the other.
%By precisely engineering optical nanostructures, one can achieve similarly strong transmission selectivity with respect to arbitrary elliptical light polarizations \cite{Wang2021}.

CD, being an optical signature of geometric chirality, should not be confused with similar effects inherent to achiral objects. For example, metasurfaces having \emph{at most} two-fold rotational symmetry around a vertical axis can exhibit unequal RCP-to-LCP and LCP-to-RCP transmission described by the so-called circular conversion dichroism: $\textrm{CCD} = (|t_{RL}|^2 - |t_{LR}|^2)/(|t_{RL}|^2 + |t_{LR}|^2)$.
Although the co-polarized transmittances are bound to be equal for achiral structures, such metasurfaces may exhibit unequal total power transmission upon illumination with RCP and LCP waves: ${\left|t_{RR}\right|}^2+{\left|t_{LR}\right|}^2 \neq {\left|t_{LL}\right|}^2+{\left|t_{RL}\right|}^2$ \cite{Fedotov2006}.
Violating high-order rotational symmetry is crucial for designing handedness-preserving mirrors \cite{Plum2015,Semnani2020}, which are an essential building block for constructing chiral polaritons.

One can also introduce the notion of a chiral source of \elm field.
An elementary chiral source is a combination of parallel electric $\bp$ and magnetic $\mathbf{m}$ point dipoles with a $\pm \pi/2$ phase difference, Fig. \ref{fig2}(c):
\begin{equation}
    \mathbf{m} = - i \xi c \mathbf{p},
    \label{Eq_10}
\end{equation}
where $\xi$ is a real-valued parameter. $\xi = \pm 1$ defines an ideal LH ('+1') and RH ('-1') chiral source, respectively, that emits only fields of the corresponding handedness \cite{zambrana2016tailoring,Fernandez-Corbaton2016}.

This notion of chiral field sources naturally translates to quantum emitters (QEs). A chiral QE is one, whose transition between the ground and excited levels is quantified by a combination of collinear electric and magnetic dipole moments satisfying Eq. \ref{Eq_10} \cite{Klimov2012}.
In realistic QEs whose transitions do possess electric and magnetic dipole moments at the same time, they are usually not aligned ideally as required by Eq. \ref{Eq_10}, but possess a perpendicular component of the magnetic dipole moment $\mathbf{m} \bot \bp$. This component can be associated with the so called Omega-type bi-anisotropic response of the medium \cite{Lindell2018}.

%\rev{[Let's talk about Lorentz models for eps and kappa.]}
Knowledge of the transition dipole moments of a particular QE allows one to describe an ensemble of those in terms of macroscopic permittivity and Pasteur parameter, as in Eq. \ref{Eq_5}.
Although often assumed non-dispersive, correct frequency dispersion of Pasteur parameter (as well as permittivity) will be crucial for description of strong coupling between quantum emitters and \elm field.
As long as a single electronic transition of the QE is involved, the permittivity and the Pasteur parameter of a macroscopic ensemble of such EQs with volume density $\rho$ can be described by the Lorentz model \cite{Condon1937}:
\begin{align}
    \eps(\omega) = \eps_{\infty} + f \frac{\omega_P^2} {\omega_0^2-\omega^2 - i \gamma \omega},\ \
    \kappa (\omega) = g \frac{\omega_P^2} {\omega_0^2-\omega^2 - i \gamma \omega},
\end{align}
where $\omega_P^2 = \rho e^2 / (3 \eps_0 m)$ is the plasma frequency of the transition ($1/3$ accounts for  isotropic orientation of QEs), $e$ and $m$ are the electron mass and charge, respectively, $f = 2m\omega_0 |\mathbf{p}|^2 / (e^2 \hbar)$ and $g = 2m\omega \Im[\mathbf{p}\cdot\mathbf{m}^*] /(e^2\hbar c)$ are the electric and magneto-electric oscillator strengths of the chiral transition, $\omega_0$ and $\gamma$ are the transition frequency and linewidth, respectively.

Accounting for light-matter interactions determined by further higher-order multipoles of molecular transitions predictably expands the palette of chiral optical effects especially arising when structured light beams are involved. Although it is hard to access the higher-order multipoles by plane light waves, tightly focused optical vortices with strongly inhomogeneous field distributions provide a propper new set of tools. Intriguing effects are observed, for example, as chiral molecular electric quadrupoles asymmetrically interact with optical vortices of different handedness giving rise to the so called  helical dichroism.\cite{Brullot2016}

\subsection{Chirality versus spin of light}

In quantum electrodynamics, the fundamental importance of chiral light forms becomes obvious already on the very basic level of analysis of intrinsic symmetry of free space photons \cite{Berestetskii2012}. For a particle with mass, such symmetry can be conveniently analyzed in its rest frame. As rest frames are generally nonexistent for photons, one has to consider the space transformations preserving the photon propagation direction.
%, i.e.,  arbitrary rotations about this direction and translations along it. 
Circularly polarized plane waves are natural eigenstates of such coordinate transforms.
% they simply acquire certain phase factors upon these rotations and translations. 
The absence of the rest frame, on the other hand, does not allow for a proper introduction of a photon \emph{spin}, which otherwise appears as an intrinsic angular momentum of a resting particle with mass. 
Considering photon parity with respect to the coordinate inversion, however, one recognizes the necessity to introduce a close analog of spin describing the intrinsic angular momentum of a photon -- its spin angular momentum (SAM). 
%While free-space photons can be shaped in space in various states with the orbital angular momentum (OAM) taking arbitrary integer values, the vectorial character of their wave functions requires introducing SAM which also contributes to the total angular momentum. 

%Moreover, although OAM and SAM are not separately conserved for a free relativistic particle, one quickly understands that the projection of photon SAM on the propagation direction is conserved. Indeed, whereas the total angular momentum (a sum of OAM and SAM) in conserved along with its projection on any direction, OAM is by definition orthogonal to the propagation direction. Therefore, the total angular momentum projection on this direction reduces to that of SAM and this particular component of photon SAM--photon helicity--appears to be a very useful conserved photon characteristic.

In terms of the classical electrodynamics, SAM of an \elm field can be treated as a continuous field variable.
From this standpoint, the SAM density is a vector quantity, which for a monochromatic \elm field $\{\bE(\br), \bH  (\br)\}$ in free space reads: \cite{bliokh2014extraordinary}:
\begin{equation}
    \bs = \frac{1}{4\omega }\Im \left[\eps_0{\bE ^*}\times \bE  + \mu_0 {\bH ^*}\times \bH \right].
    \label{Eq_6_spin_4}
\end{equation} 
A trivial situation wherein an \elm field possesses SAM density is presented by an RCP or LCP plane wave. In this case, the SAM density is parallel or anti-parallel to the wave vector depending on the handedness of the wave.

%{\color{olive} \em Honestly, I consider the concept of vectorial SAM arguable (see the commented Landau's arguments above). On the other hand, do we encounter evanescent waves in chiral polaritons? Do we need all this here?}

Another far less obvious situation when an \elm field features a spin is presented by evanescent plane waves. Consider a monochromatic transverse-magnetic evanescent wave propagating in free space along the $x$ axis, and attenuating exponentially in the $z$ direction: $\bH \propto \uy e^{i k_x x - \kappa_z z}$, $\bE \propto \left(- i\kappa_z \ux + k_x \uz \right) e^{i k_x x - \kappa_z z}$,
where the wave vector components satisfy $k_x^2 - \kappa_z^2 = (\omega/c)^2$. After simple algebra we find that such a wave also carries a non-zero SAM density component, which is perpendicular to the propagation and the attenuation direction \cite{bliokh2014extraordinary}:
\begin{equation}
    \bs_T \propto \Re\ \bk  \times \Im \ \bk.
    \label{Eq_6_spin_6}
\end{equation} 
%What is intriguing about this results is the presence of a $y$-component of SAM in the wave propagating in the x-direction.
This component of SAM arises due to the elliptical polarization of the electric field in the propagation plane, Fig. 4(a), and is known as the \textit{transverse spin angular momentum} \cite{bliokh2014extraordinary, bliokh2015spin}. 

\begin{figure*}[hbt!]
\includegraphics[width=.8 \textwidth]{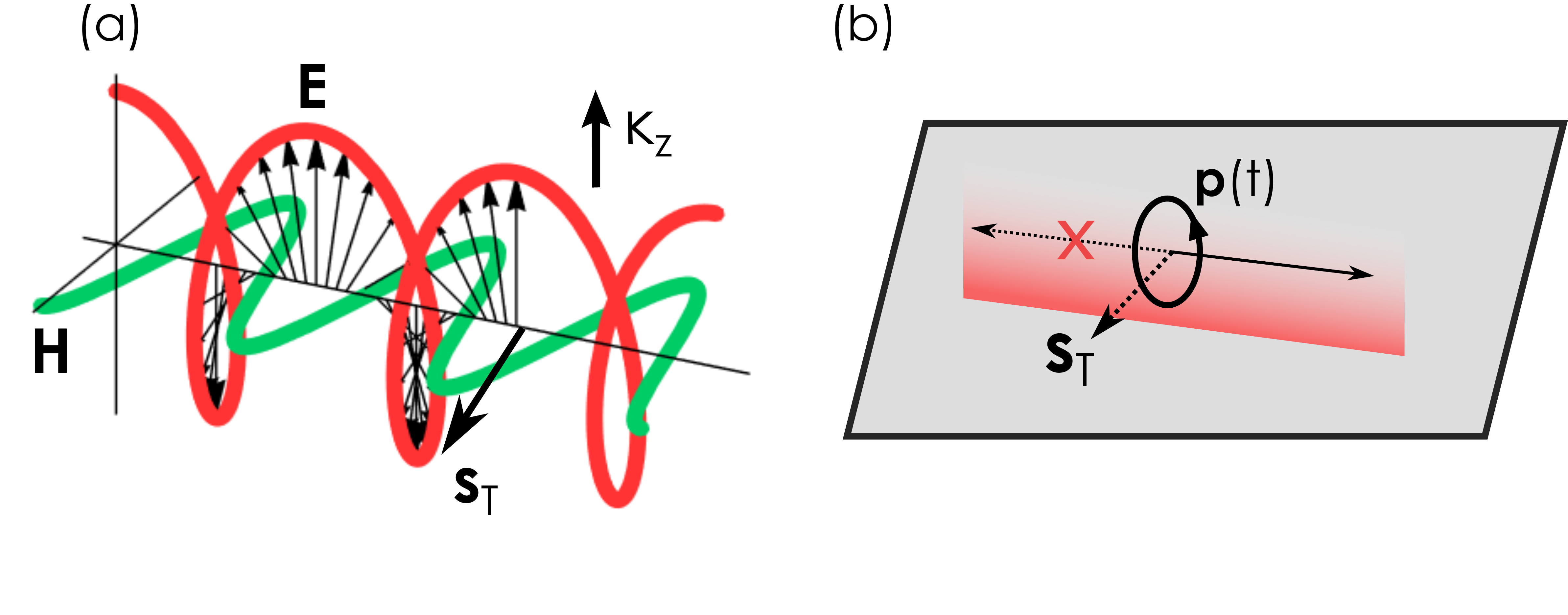}
\caption{\textbf{Relation between chirality and spin.} (a)  Illustration of the spatial distribution of the electric and magnetic fields in a TM-polarized evanescent wave propagating along the $x$ axis and exponentially decaying along the $z$ axis. Elliptical polarization of the electric field produces a transverse component of the spin angular momentum perpendicular to the linear momentum of the wave (b) Spin-momentum locking in evanescent waves. An elliptically polarized electric dipole source placed near a waveguiding surface features directional coupling to guided waves due to the spin-orbit coupling.}
\label{fig4}
\end{figure*}

This property of evanescent fields leads to a fundamental connection between light's SAM density and linear momentum near planar surfaces and enables unidirectional excitation of waveguide modes in various nanophotonic platforms \cite{rodriguez2013near, bliokh2015spin, lodahl2017chiral}.
Consider a TM-polarized bound mode supported by an interface (such as a surface wave propagating along a metallic surface, or a waveguide mode of a dielectric film). 
An electric dipole with an appropriate elliptical polarization placed near a waveguiding surface will produce an \elm field that will couple preferentially to those guided modes, whose transverse SAM density matches that of the circular-polarized source, Fig. 4(b). 
An analogous spin-momentum locking arises for TE-polarized bound modes \cite{wang2018magnetic}. 

%The \elm field of this mode is described by an evanescent wave, whose electric field has an elliptical polarization. This elliptical polarization leads to a transverse spin angular momentum component $\bs_T$, which is perpendicular both to the interface (that matches the evanescent field attenuation direction) and the field's linear momentum. Thus, the direction of this spin angular momentum gets locked to the guided mode propagation direction \cite{van2016universal}. 

It is crucial to emphasize the similarities and the differences between the SAM density and the chirality of light.
SAM density is a time-odd vector, while chirality density is a time-even pseudo-scalar.
To put this into perspective, one may consider a circularly-polarized homogeneous wave. The SAM density in such a wave is collinear with the momentum, and time reversal leaves its handedness unchanged, but flips the SAM density.
A lot of excellent works dealing with this mechanism of unidirectional excitation have used terms "chiral coupling", "chiral emission", and alike (see, e.g., refs. \cite{petersen2014chiral, lodahl2017chiral, chervy2018room, sollner2015deterministic, guo2019routing}), while the studied systems had little to do with true chirality of light. Chirality involves chiral sources represented by a combination of collinear electric and magnetic dipoles, as opposed to circularly polarized electric dipole engaged in the effect of spin-orbit coupling.
%In fact, the connection between chirality and spin arises at a different level: the two quantities assemble a form of a conservation law, where the SAM density describes the flux of chirality density \cite{Bliokh2011}.

\section{Constructing chiral polaritons}

Now we are in a position to discuss the notion of a chiral polariton, by which we are going to refer to the coupled state of an interacting light-matter system, where both the optical mode \emph{and} the material excitation are chiral.
Maxwell's equations in free space admit solutions in the form of chiral photons, so one could already examine chiral polaritons emerging in this situation. However, from the practical standpoint, we are usually interested in strong interaction with confined optical modes. %in this case both handednesses coexist at the same time; furthermore
Thus, it would seem natural to load an ordinary resonator -- such as a \fp cavity mode -- with chiral matter and expect excitation of chiral polaritons upon excitation of such a system with a circularly polarized \elm field.
This kind of system has been realized in ref. \cite{Baranov2020CD} by coupling chiral metallic meta-atoms to the normal-incidence resonance of a \fp cavity. Similar setup was realized in ref. \cite{guo2021optical} by coupling chiral organic emitters to a non-chiral surface plasmon mode of a metallic film.

\begin{figure*}[hbt!]
\includegraphics[width=.9\textwidth]{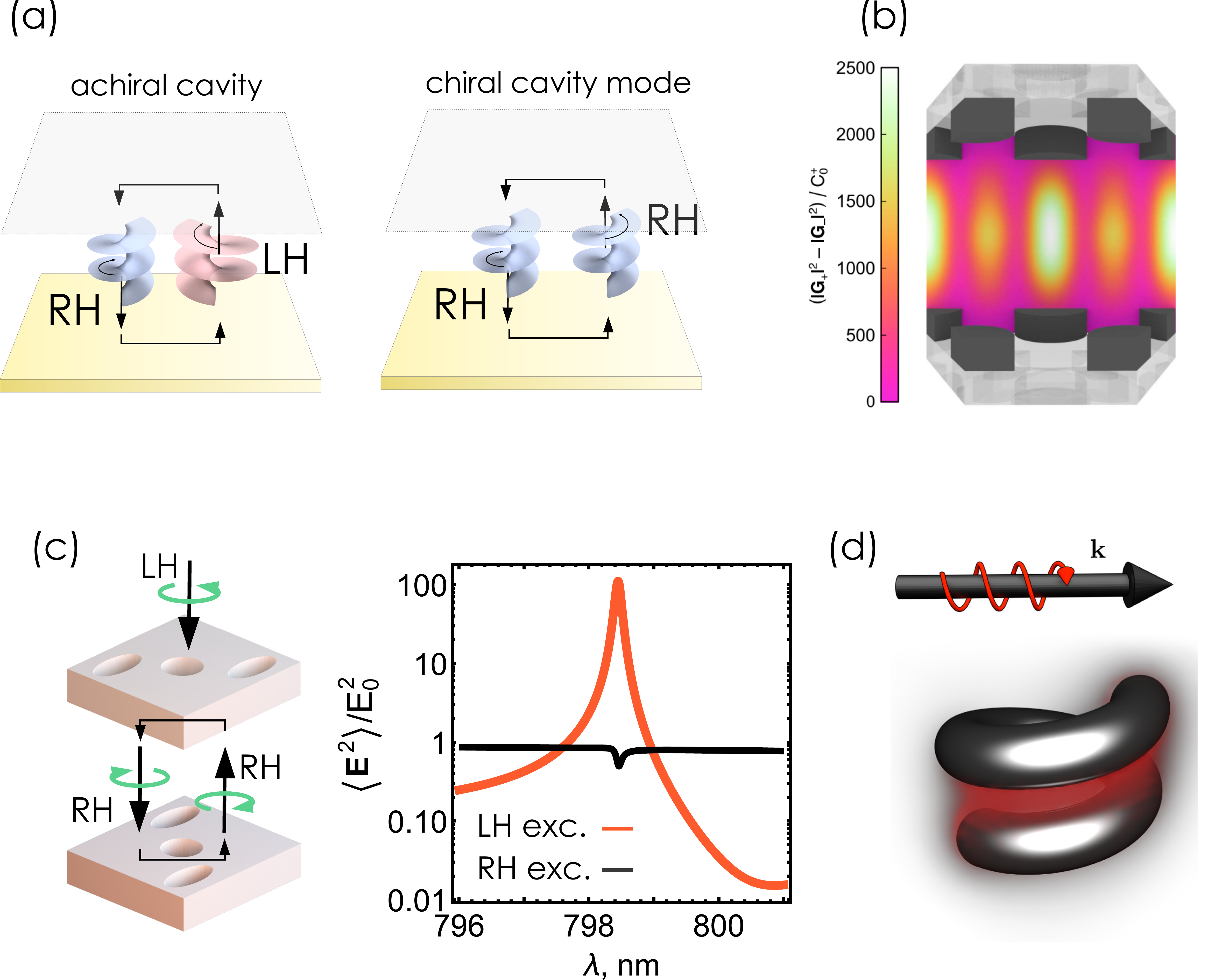}
\caption{\textbf{Handedness-preserving optical cavities.} (a) The difference between an ordinary \fp cavity, and a handedness-preserving planar cavity. An ordinary \fp resonator supports standing waves with linear polarization. A hypothetical handedness-preserving cavity, on the contrary, supports a chiral standing wave with a linear polarization at every point in space. (b) Sketch of the handedness-preserving cavity composed of two dielectric photonic crystal mirrors. The color shows the optical chirality density of the resulting standing wave. (c) Left: single-handedness chiral cavity realized by stacking two dielectric low-symmetry mirrors. Right: simulated electric field enhancement spectra. (d) A compact maximally chiral optical cavity represented by a metallic helix producing a resonant response only to one incident handedness and being nearly transparent to the other handedness.
Panel (b) adapted with permission from ref. \cite{Feis2020}.
Parts of panel (c) adapted from ref. \cite{Voronin2022}.
Panel (d) adapted with permission from ref. \cite{garcia2022toward}.
}
\label{fig5}
\end{figure*}

%To exemplify the problem, consider a circularly polarized standing wave inside a \fp resonator consisting of two isotropic homogeneous metallic mirrors. The field continuity at the surface of the mirror require that both waves are described by the same complex polarization vector ${\left(1, \pm i \lambda,0\right)}^T$
The simple intuition outlined above faces a fundamental obstacle.
To exemplify the problem, consider a perfect electric mirror illuminated at normal incidence by an LCP or an RCP plane wave $\bE_{inc}\left( \br \right) \propto {\left(1, \pm i \lambda,0\right)}^T  e^{ikz}$, $\lambda = \pm 1$. 
The field continuity at the surface of the mirror requires that the reflected wave is described by the same complex polarization ${\left(1, \pm i \lambda,0\right)}^T$. 
Thus, reflection of the circularly polarized wave at a normal by a metallic mirror preserves its SAM density, but reverses its wave vector, which in the end flips the handedness of the reflected wave.
This handedness flipping of the wave travelling between two mirrors creates a standing wave with zero chirality density, Fig. \ref{fig5}(a). As a result, the quasinormal modes (QNMs) of a \fp cavity at normal incidence do not possess any handedness. This property of circularly polarized standing waves imposes severe constraints on the behavior of polaritonic systems.
In particular, this helicity flipping ultimately limits the magnitude of CD that can be observed with a chiral film strongly coupled to an ordinary \fp cavity despite the resonant enhancement of the \elm field inside the cavity \cite{Baranov2020CD}.

The problem of handedness flipping poses a quest for optical cavities supporting QNMs of well-defined handedness, whose performance is schematically illustrated in Fig. \ref{fig5}(a) \cite{hubener2021engineering}.
A natural question then arises: what design rules does one need to follow in order to engineer a cavity that preserves the handedness of \elm field (upon scattering)? This turns out to be a more fundamental issue which we address below.

\subsection{Handedness-preserving cavities}

Handedness of light may be converted in different ways upon interaction with scatterers, even non-chiral ones. 
As we showed above, any field distribution satisfying Maxwell's equations can be expanded into two components of well defined handedness $\bG_{\pm }$.
Imagine a scenario when an incident field ${\bE }_{inc}( \br) $ of a certain handedness (for example, a purely LH field) interacts with an arbitrary scatterer, such as a metasurface or a compact nanoparticle. For this field only the $\bG_+$ component is different from zero. The scattered field, on the other hand, would generally contain both handedness components: $\bG^{scat}_+ \left( \br \right) \ne 0, \bG^{scat}_- \left( \br \right) \ne 0$.
%In the considered case of a normally incident plane wave reflection by a homogeneous metallic mirror a LH normally incident light is converted completely to RH light.

Some cavities, such as plasmonic and dielectric nanoantennas, exhibit spatial regions of enhanced chirality density even upon excitation with linearly polarized light \cite{Davis2013,Schaferling2012,gautier2022planar}, but QNMs of these systems still do not possess a well-defined pure handedness.
In this context, the concept of duality \cite{Fernandez-Corbaton2013} plays an important role.
Using the framework of Riemann-Silberstein vectors one can show that a piece-wise homogeneous and isotropic system preserves the handedness of any incident \elm field at a frequency $\omega$ if and only if all the materials have a constant impedance across all media constituting the system, including the environment \cite{Fernandez-Corbaton2013}:
\begin{equation}
    \frac{\eps\left(\br, \omega \right) }{\mu (\br, \omega)}
    = \frac{\eps_{env}\left(\br, \omega \right) }{\mu_{env} (\br, \omega)} 
    = \rm const \ \ \forall \br.
    \label{Eq_12}
\end{equation}
Structures that satisfy this criterion are referred to as \textit{dual}. Correspondingly, if the structure is surrounded by free space, Eq. \ref{Eq_12} requires equal permittivity and permeability of the resonator, $\eps(\br, \omega) = \mu (\br , \omega)$.

As one can see, duality is a very stringent requirement that calls for magnetic materials, $\mu \ne 1$. Since they are very rare in the visible range, this motivates the search for non-magnetic handedness-preserving structures. % that preserve the handedness at least for certain excitation wavelengths and directions. Recently, there has been a considerable progress in the development of such nanostructures.
Feis \emph{et al.} \cite{Feis2020} demonstrated a handedness-preserving cavity formed by a pair of periodic dielectric mirrors, Fig. \ref{fig5}(b).
Each mirror operates by virtue of first-order diffraction into large transverse momentum ($k_{\parallel}$) modes. 
Thanks to the presence of an inversion center, the system itself is achiral, meaning it supports both the LH and the RH QNMs at the same time. 
Voronin et al. pushed this idea further and studied a \emph{single-handedness} chiral optical cavity \cite{Voronin2022}.
%An additional requirement that one could impose is that the structure supports a QNM of a certain handedness only, but not that of the opposite one (at least in a given spectral range).  The idea of such a resonator has been outlined in ref. \cite{plum2015chiral}. %although the behavior of the resulting structure has not been addressed.
By utilizing low-symmetry dielectric mirrors adopted from ref. \cite{Semnani2020}, they designed a \fp cavity that supports a QNM of a well-defined handedness, and does not support QNMs of the opposite handedness in the relevant wavelength range, Fig. \ref{fig5}(c). 
%Resonant excitation of the cavity with light of appropriate handedness induces a chiral standing wave with a uniform chirality density, while the light of opposite handedness does not cause any resonant effects.

In addition to planar infinitely extended resonator, there has been a progress in designing compact nanostructures with chiral optical resonances.
Garcia et al. experimentally demonstrated what is called a maximally chiral object \cite{Fernandez-Corbaton2016} of a finite size operating in the IR \cite{garcia2022toward}. Their structure is a helix-shaped metallic nanoresonator, Fig. \ref{fig5}(d), that is nearly transparent for incident light of a certain handedness, while excitation with the opposite handedness produces resonant extinction. 
Any maximally chiral object is necessarily dual \cite{Fernandez-Corbaton2016}, implying that the excited QNMs should have a well defined handedness.
The interior of the resulting nanocavity can be loaded with an excitonic material similarly to ref. \cite{stamatopoulou2022reconfigurable}, which studied theoretically a similar plasmonic nanostructure interfaced with an achiral excitonic material.

\subsection{A theoretical model for chiral polaritonics}

%\Cadd{I shortened it a bit, lets talk another time for more refinement.}
%\rev{[DGB: I personally would avoid presenting the full Hamiltonian and the field expansion: these are technical details that fall outside the scope of the perspective. If I wanted to keep the H, I'd try presenting it in a more schematic way, like H is matter + light + interaction + P2 + M2, the interaction is this, and we are unsure about the M2 because of this and that, but so far the result is this. Let's say we want the same, or slightly more advanced level of difficulty as in Eqs. 1 and 2. Can we do that?]}

The availability of a handedness-preserving cavity is only one part of the story, next we have to understand how to utilize it. Exploring the possibilities of chiral polaritonics is no small endeavour and will demand ample work on both theoretical and experimental side. Intuition, often generated by simple models as illustrated in this section, serves as an important catalyst. We refer the interested reader to Ref.~\cite{schafer2022chiral} for a detailed derivation.

\begin{figure}[h!]
\includegraphics[width=.5\textwidth]{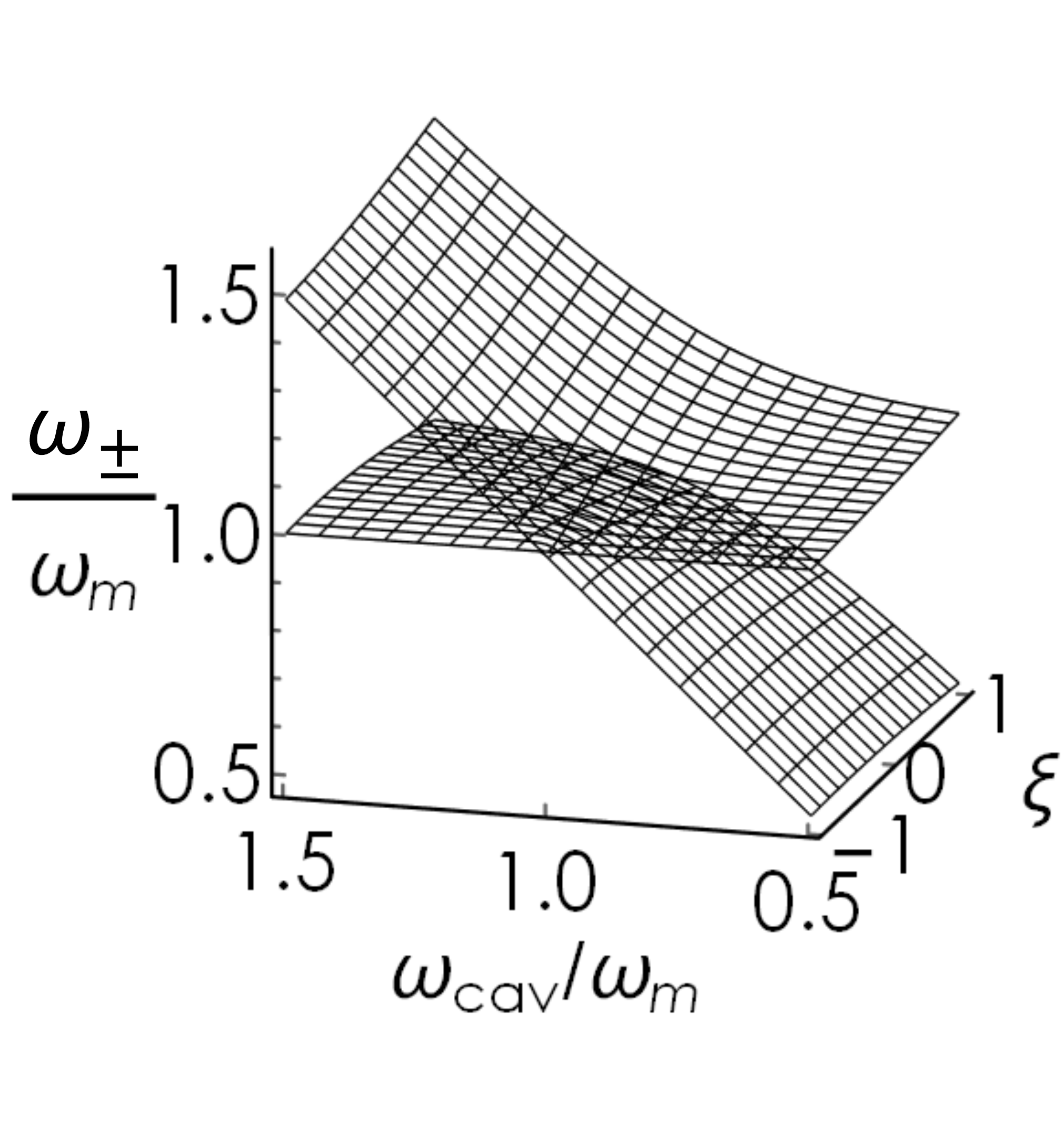}
\caption{Spectrum of polaritonic eigenvalues $ \omega_{\pm}$ of the chiral Hopfield model with a left-handed cavity mode as a function of the cavity frequency $\omega_{cav}$ in relation to the matter excitation $\omega_m$ and chiral factor $\xi$. The eigenvalues are calculated with typical values for dye molecules (see Ref.~\cite{schafer2022chiral} for details).
}
\label{fig:hopfieldscan}
\end{figure}

The starting point for all theoretical descriptions is the non-relativistic minimal-coupling Hamiltonian.
Since molecules are usually small in relation to the wavelength of the confined modes it is beneficial to apply the Power-Zienau-Wooley transformation and retain only the electric and magnetic dipolar terms as well as electric quadrupoles \cite{babiker1974generalization,andrews2018perspective}. The lowest-order Hamiltonian for chiral interaction takes therefore the form 
\begin{align}
\label{eq:Hlm}
    \hat{H} &= \hat{H}_M + \hat{H}_L -\frac{1}{\eps_0} \sum_n^N \hat{\boldsymbol\mu}_n \cdot \hat{\bD}_\perp(\br_n) - \sum_n^N \hat{\textbf{m}}_n \cdot \hat{\bB}(\br_n)
    + \hat{H}_{LMc}
    %- \frac{1}{\eps_0} \sum_n \sum_{a,b\in \{x,y,z\}} \hat{Q}_{ab,n} \nabla_{a,n} \hat{D}_{\perp,b}(\br_n)\notag\\ 
    %&+ \frac{1}{8m} \sum_{n,i} (q_{n,i}\hat{\br}_{n,i} \times \hat{\bB}(\br_n))^2 + \frac{1}{2\eps_0} \int d^3 \br \hat{\bP}_\perp^2,    
\end{align}
%\begin{align}
%\label{eq:Hlm}
%    \hat{H}_{M} &= \sum_{n=1}^{N} \sum_{i=1}^{N_M} \frac{1}{2m_i} \hat{\bp}_{i,n}^2 + \sum_{n=1}^N \hat{V}^n_{\parallel}(\hat{\br}) + \sum_{n\neq n'}^{N} \hat{V}^{n,n'}_{\parallel}(\hat{\br}), \notag \\ 
%    \hat{H}_{L} &= \frac{1}{2} \int d^3 \br \left[ \frac{\hat{\bD}_\perp^2(\br)}{\eps_0} + \eps_0 c^2 \hat{\bB}^2(\br) \right]. \notag \\
%    \hat{H}_{LM} &= -\frac{1}{\eps_0} \sum_n \hat{\boldsymbol\mu}_n \cdot \hat{\bD}_\perp(\br_n) - \sum_n\hat{\textbf{m}}_n \cdot \hat{\bB}(\br_n)\\ 
%    &- \frac{1}{\eps_0} \sum_n \sum_{a,b\in \{x,y,z\}} \hat{Q}_{ab,n} \nabla_{a,n} \hat{D}_{\perp,b}(\br_n) \notag\\
%    &+ \frac{1}{8m} \sum_{n,i} (q_{n,i}\hat{\br}_{n,i} \times \hat{\bB}(\br_n))^2 + \frac{1}{2\eps_0} \int d^3 \br \hat{\bP}_\perp^2,    \notag
%\end{align}
with the electric and magnetic transition-dipole moments $\hat{\boldsymbol\mu}_n,~\hat{\textbf{m}}_n$ related by the transition elements $\textbf{m}_n^{01} = -ic\xi \boldsymbol\mu_n^{01}$, displacement $\hat{\bD}_\perp(\br_n)$ and magnetic  $\hat{\bB}(\br_n)$ field. The corrections collected in $\hat{H}_{LMc}$ are essential for gauge-invariance, realistic descriptions and large interaction strength \cite{schafer2022chiral}.
%where the last line collects self-interaction terms that are important for gauge-invariance and the stability of the combined system. Eq.~\eqref{eq:Hlm} represents a possible starting point for all levels of theory, may it be \textit{ab initio} or model. 
Importantly, the dipolar moments are defined with respect to the center of mass of each individual molecule. %In a next step we have to specify how the quantized standing chiral fields look inside a handedness preserving cavity. Following Ref.~\cite{schafer2022chiral} , one identifies as simplest realization
%\begin{align*}
%    \hat{\bD}^{\lambda}_\perp(\br) &= - \sqrt{\frac{\eps_0}{V}} \tilde{\eps}^{\lambda}_{k}(z) \hat{p}_{k,\lambda},\\
%    \hat{\bB}_{\lambda}(\br) &= \sqrt{k^2/\eps_0 V} \lambda \tilde{\eps}^{\lambda}_{k}(z) \hat{q}_{k,\lambda}
%\end{align*}
%where $\tilde{\eps}^{\lambda}_{k}(z) =  \left(\cos(k z), - \lambda \sin(k z), 0 \right)^T$, $\lambda$ being the handedness of the chiral field and $\hat{q},~\hat{p}$ the canonical position and momenta. Given those ingredients and following Ref.~\cite{schafer2022chiral} , 
Following Ref.~\cite{schafer2022chiral}, we can obtain an intuitive solution for the interaction between N strongly simplified chiral emitters and the standing chiral field. %The combined system features an avoided crossing
%The transition energies of the coupled system takes the form:
In a \textit{simplified} form, the analytic solution 
\begin{align}
    \label{eq:chybrid}
    \omega_{\pm} \approx \frac{1}{\sqrt{2}}\sqrt{ \omega_{cav}^2 + \omega_0^2 + 8 \xi\lambda g^2 \pm \sqrt{[\omega_{cav}^2 - \omega_0^2]^2 + 16 g^2 (\omega_{cav} + \omega_0 \xi \lambda ) (\omega_{cav} \xi \lambda + \omega_0)  } }
    %\Omega_{\pm} = \frac{1}{\sqrt{2}}\sqrt{ \bar{\omega}_k^2 + \tilde{\omega}_m^2 + 8 \tilde{\xi}\lambda N \tilde{g}^2 \pm \sqrt{[\bar{\omega}_k^2 - \tilde{\omega}_m^2]^2 + 16 N \tilde{g}^2 (\bar{\omega}_k + \tilde\omega_m \tilde\xi \lambda ) (\bar{\omega}_k \tilde\xi \lambda + \tilde\omega_m)  } }
\end{align}
provides an intuitive way to understand the interplay of the handedness of field $\lambda$ and molecules $\xi$. The polaritonic eigenvalues $E_{\pm}  = \hbar \omega_{\pm} + E_{vac}$, where $E_{vac} = \hbar (\omega_{+} + \omega_{-})/2$ is the ground state energy of the coupled chiral system, will exhibit an avoided crossing $\propto g (1+\lambda \xi) $.
Fig.~\ref{fig:hopfieldscan} illustrates this handedness depending hybridization that gives rise to a plethora of possible applications.
%The hybridization between the renormalized light and matter excitations $\bar{\omega}_k,~\tilde{\omega}_m$ scales as $\sqrt{N}\tilde{g} (1+\lambda\tilde\xi)$ where $\tilde{g}$ is the renormalized fundamental coupling strength and $\tilde\xi$ the effective emitter chirality. 

The handedness preserving mirrors are therefore able to lift the degeneracy between same ($\lambda \xi > 0$) and opposite ($\lambda \xi < 0$) combinations -- we obtain non-intrusive control over the enantiomers. Furthermore, this extends also to the correlated ground-state of the combined system. While it sounds like a dream come true, it is important to realize that the degree of chirality is often weak and the energetic difference between the enantiomers therefore small. Nevertheless, chiral polaritionics suggests a plethora of possible applications, some will demand better engineered fields, others might work out-of-the-box. In the following, we will discuss some of those applications, debating short-term challenges and long-term prospects. Furthermore, we briefly discuss which theoretical approaches might be suited and how existing techniques can be extended.

\section{Prospects and Challenges in Chiral Polaritonics}
%\Cadd{The following section is now starting the 'perspective' part. It would be maybe nice to have an overall 50\%-50\% split between the 'review/basics' sections and the 'perspective/outlook' sections.}\\
%\Cadd{This is a first draft/outline of what one could write here, let me know what you think.}\\

Chirality is a long standing and widely impactful concept in natural science. If we add sugar to our coffee, enjoy a good meal, or are in the need of medication - chirality has a defining influence at every point of our life and even the existence of life, as we know it, itself. Understanding and controlling the chiral state of a substance is thus pivotal. Early work in chiral catalysis was awarded in 2001 with the Nobel price in chemistry \cite{NobelChem}. In the following, we will illustrate a selected set of promising directions, discuss associated challenges and highlight recent development in the theoretical description of chiral polaritonics.

\subsection{Anticipated novel phenomena}
\label{sec:perspective}

%The class of chiral optical cavities opens a number of intriguing opportunities for studies of interaction between chiral QEs and \elm field, both in the weak and strong coupling regimes

\paragraph{Controlling spontaneous emission:}

\begin{figure*}[hbt!]
\includegraphics[width=.8\textwidth]{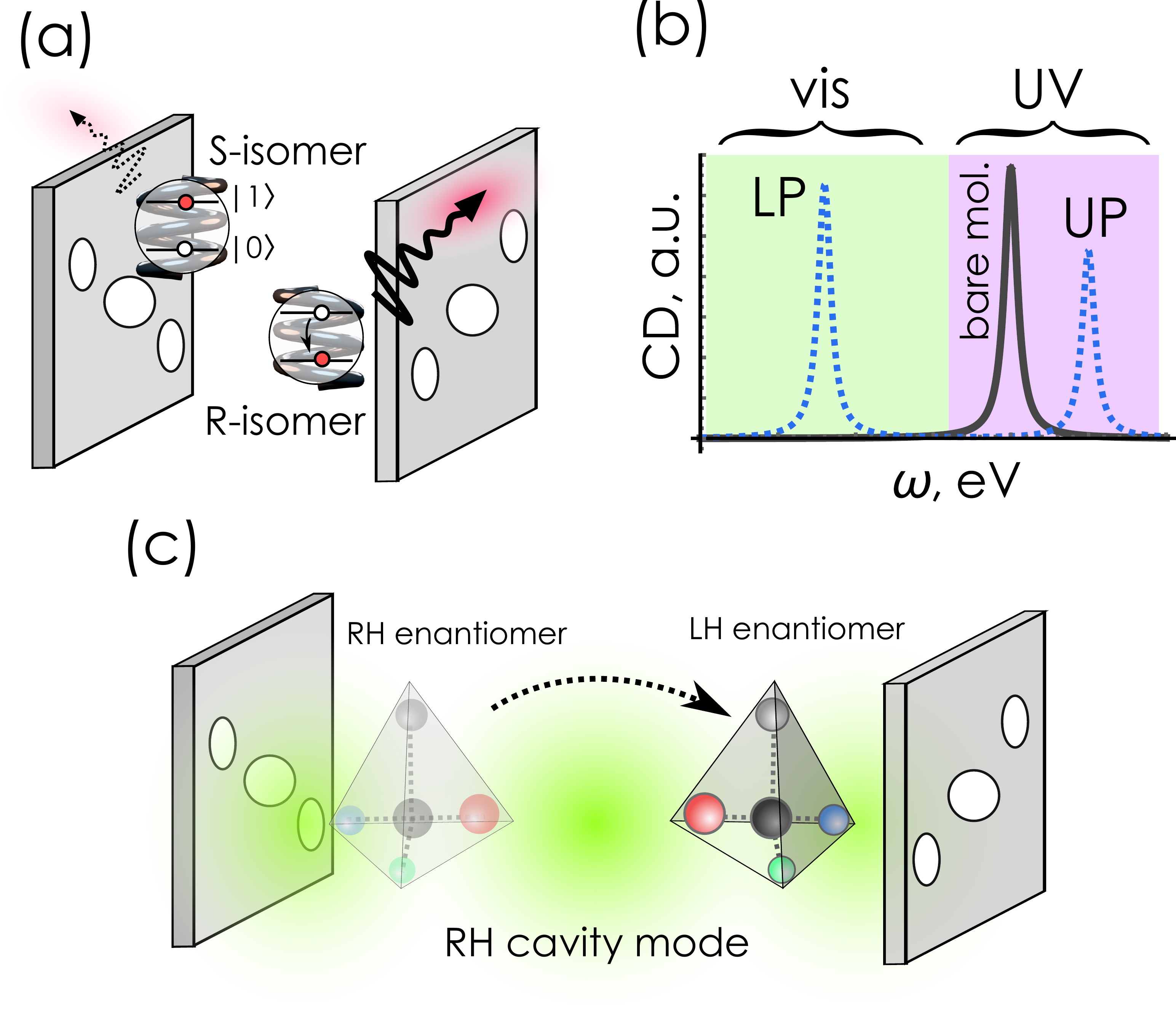}
\caption{\textbf{Potential implications of chiral strong coupling.} (a) Selective de-excitation of chiral molecules. Two opposite enantiomers of a chiral molecule coupled with an asymmetric single-handedness optical cavity. Asymmetric coupling strength leads to unequal spontaneous decay rates from the excited states of two enantiomers. (b) Red-shifting chiral transitions by strong coupling. A single chiral electronic transition of a molecule located in the UV range. Strongly coupling this molecule with a photonic mode causes of the the lower polariton (LP) to red-shift towards the visible range while the upper polariton (UP) is shifted deeper into the UV domain. (c) Steering the equilibrium of an ensemble of racemic mixture by ultra-strong coupling of chiral emitters to a single-handedness optical cavity.} 
\label{fig7}
\end{figure*}

In the weak coupling regime, single-handedness cavities could enable differential spontaneous emission rates for left and right-handed enantiomers of chiral sources.
Yoo and Park introduced the notion of chiral Purcell factor back in 2015 \cite{yoo2015chiral}; however, their description can be questioned because the authors assumed that LH and RH enantiomers of a given chiral source would show unequal spontaneous emission rates in free space, while in fact they must be equal (the empty environment of the QE is symmetric with respect to inversion).
Voronin et al. studied the radiation of chiral point sources coupled to their single-handedness cavities and observed a strong asymmetry in the emission intensities, suggesting that the fluctuating \elm field within a single-handedness cavity acquires a strong chiral character \cite{Voronin2022}. Such an observation follows intuitively from the handedness-dependent mode density in a chiral environment that gives rise to handedness-dependent Purcell factors.
Thus, electronically excited states of opposite enantiomers of a chiral molecule would acquire different lifetimes, Fig. \ref{fig7}(b), which could be interesting in the context of selective de-excitation of chiral molecules \cite{Klimov2012}.
Recent studies demonstrate intrinsic asymmetry of photoluminescence from chiral excitons hosted in J-aggregates \cite{li2022strong}; single-handedness cavities may further enhance this emission asymmetry.

%\Cadd{[Flip sections or labels in FIG 7? At the moment thi section is first and refers to (b) while the next refers to (a).]}

\paragraph{Red-shifting chiral transitions:}

Either symmetric two-handedness or asymmetric single-handedness chiral cavities can be utilized for engineering of chiral polaritons by loading them with resonant Pasteur media, which could have promising implications for chiral sensing.
Indeed, chiral electronic transitions of the majority of bio-molecules are located in the UV range, where spectroscopic measurements are challenging.
The two polaritons of a strongly coupled system are expected to experience an anti-crossing, wherein the lower polariton shifts towards the visible range, Fig. \ref{fig7}(a), where measuring CD and characterizing the enantiomeric state of a racemic would be much easier.
Recent theoretical efforts are beginning to explore this strategy. For example, Feis et al. and Beutel et al. studied the response of a symmetric handedness-preserving cavity loaded with a non-resonant Pasteur medium and shown enhancement of CD \cite{Feis2020,Beutel2021}, while Mauro et al. examined the behavior of a single-handedness cavity loaded with \emph{resonant} chiral medium \cite{mauro2022chiral}.
%Importantly, although non-chiral \fp cavities could lead to the formation of polaritons with chiral emitters, they are not suited for this purpose due to the non-chiral nature of their QNMs, which limits the observed CD magnitude \cite{Baranov2020CD}.

%\Cadd{[we should cite and discuss here the recent work by Feist and Avriller arXiv:2209.00402. The more vivid we paint the picture of the field the better for the field/perspective, so lets add everything we find that is already discussed.]}

\paragraph{Leveraging chiral vacua:}

The chiral nature of vacuum fluctuations in single-handedness cavities might enable unusual vacuum energy landscapes.
A single harmonic oscillator possesses a vacuum (a.k.a. zero-point) energy of $U_{vac} = \hbar \omega/2$. In a polaritonic system made up of two interacting harmonic excitations the vacuum energy therefore is $U_{vac} = \hbar(\omega_{+} + \omega_{-})/2$, where $\omega_{\pm}$ are the two polaritonic transition energies.\cite{ciuti2005quantum,Baranov2020USC}
Such changes in the correlated ground state will differ for each enantiomer and increase with the molecular concentration as $\sqrt{N}$.\cite{schafer2022chiral,riso2022strong} 
%Solving non-chiral Hopfield Hamiltonian shows that the vacuum energy of a strongly coupled polaritonic system scales quadratically with the coupling strength, $U_{vac} \propto |g|^2$, ref. \cite{ciuti2005quantum,Baranov2020USC}. 
%This suggests that an ensemble of left-handed chiral emitters couples strongly to the left-handed QNM of a single-handedness cavity wil experience a greater vac energy than ...
%One can imagine a situation where an ensemble of left-handed chiral emitters couples strongly to the left-handed QNM of a single-handedness cavity; at the same time, the emitters in the opposite (right-handed) enantiomeric state are in the weak coupling regime with the field, for as long as the cavity does not support a right-handed QNM.
This hints at a possibility that chiral molecules will experience different ground-state energies inside a single-handedness cavity depending on their enantiomeric state, thus steering the molecular structure to a different equilibrium.
Irrespective the exciting possibilities, a few obstacles remain. First, molecular chirality is small, a challenge and opportunity to be revisited in the following. Second, simplified few-mode theories are well suited to describe spectral effects but can be expected to provide only a qualitative trend for the correlated ground state energy that would be predicted following Casimir. \cite{Rodriguez2007,Lambrecht2006}

\paragraph{Enhancing the Weak Chirality:}

%Here we can state the problem of small chirality factors, goal now would be to design a system that enhances that. Brings plenty of challenges (which we should explain here). Possible approaches could use plasmonic systems, Van-der-Waals structures [https://arxiv.org/pdf/2208.06249.pdf]
The parameter $\xi$ characterizing the degree of matter chirality is extremely weak for realistic molecules ($\xi \sim 10^{-4}...10^{-6}$, ref \cite{Condon1937}), which translates into a minute asymmetry of the polaritonic energy spectra between right and left enantiomers. 
One possibility to address this issue is to design a chiral cavity that would compensate for this smallness. 
The discriminating effect in an ensemble of emitters coupled to a single chiral mode scales with $\sqrt{N}g \lambda \xi$, where $\lambda$ characterizes the relationship between the electric and magnetic fields of the chiral optical mode.
This suggests that chiral molecules with $|\xi| \ll 1$ would produce a substantial energy asymmetry if they are coupled to a chiral optical mode with $\lambda \gg 1$. 
%An ideal chiral molecule coupled to the \elm field of a single-handedness cavity with $\lambda = \pm 1$ then yields $1+\tilde{\xi} \lambda = 0$ or $1+\tilde{\xi} \lambda = 2$ depending on its enantiomeric state.
This is not possible in purely transverse chiral standing waves, where the electric and magnetic field are related via $Z \bH = \pm i \bE$. However, in subwavelength cavities, where the field has a substantial longitudinal component, it does no longer hold . 
Thus, the issue could be addressed by designing a compact chiral nanocavity, as in ref. \cite{garcia2022toward}, whose quasi-normal mode is dominated by the longitudinal magnetic field, $|Z\bH| \gg |\bE|$, and maintains a chiral character.

\paragraph{Exploring autocatalysis and homochirality:}
Why do sugar (sucrose) and many other substances exist in nature in such high excess compared to their chiral images? What leads to this large imbalance between enantiomers? This simple question remains unanswered, although it marks a defining problem for the development of life.\cite{sciencechiral}
While still debated, it is conjectured that an initial unbalance between the enantiomers was amplified by autocatalytic reactions of chemical and/or biological nature \cite{blackmond2004asymmetric}. The origin of this unbalance remains however unknown, and speculations range from a pure chance due to statistical fluctuations to the parity symmetry breaking in weak interactions \cite{quack2022perspectives}. The problem is obvious, without a way to carefully adjust minute changes in the environmental conditions, it is challenging to observe and manipulate a possible mechanism.

Chiral polaritonics might provide here a unique possibility. The optical fields inside the chirality-preserving cavity undergo vacuum fluctuations that break the chiral symmetry of free space\cite{schafer2022chiral}. How strong the electromagnetic environment varies inside the cavity from the free-space conditions can be well controlled by changing size, shape and quality of the optical mirrors.
It has been demonstrated that optical cavities can be used to control anything from single atoms \cite{goy1983observation}, over chemical reactions \cite{garcia2021manipulating}, up to biomolecules and bacteria \cite{coles2017nanophotonic}.
Such a sensitive tool could now be used to fine-tune the degree to which the electromagnetic environment prefers a given handedness. Control over the chiral environment allows then control over the initial conditions of homochirality.
Furthermore, as the chiral polaritonic interaction depends on the number of collectively coupled emitters with the corresponding handedness, this effect is inherently self-catalysing. While the details of this approach remain to be investigated, chiral polaritonics could serve as a novel tool to explore homochirality.

%\paragraph{Stuff we should briefly refer to}
%\Cadd{Did we discuss/cite this somewhere: \url{https://doi.org/10.1021/acsnano.1c06959}}
%The twists and turns of chiral chemistry \cite{mackenzie2021twists}

%\paragraph{\Cadd{Other stuff that we could mention and briefly discuss:}}
%Orbital angular momentum of twisted light: chirality and optical activity \cite{forbes2021orbital}. (I think we should definitely mention the relation with twisted light, its a quite hot topic currently)
%Controlling the symmetry of inorganic ionic nanofilms with optical chirality \cite{kelly2020controlling}

%Switchable enantioseparation based on macromolecular memory of a helical polyacetylene in the solid state \cite{shimomura2014switchable}
%Complete chiral symmetry breaking of an amino acid derivative directed by circularly polarized light \cite{noorduin2009complete}
%\Cadd{Do we want to comment/discuss the time-reversal breaking cavities proposed in \cite{hubener2021engineering}? Maybe this provide a good opportunity to clarify the discussion around chirality, standing chiral waves and time-reversal symmetry in chiral and Faraday cavities. We could also just mention the perspective/ideas in the previous sections where the time-reversal symmetry is discussed.}

\subsection{Development of Theoretical Descriptions}

\begin{figure*}[hbt!]
\includegraphics[width=1\textwidth]{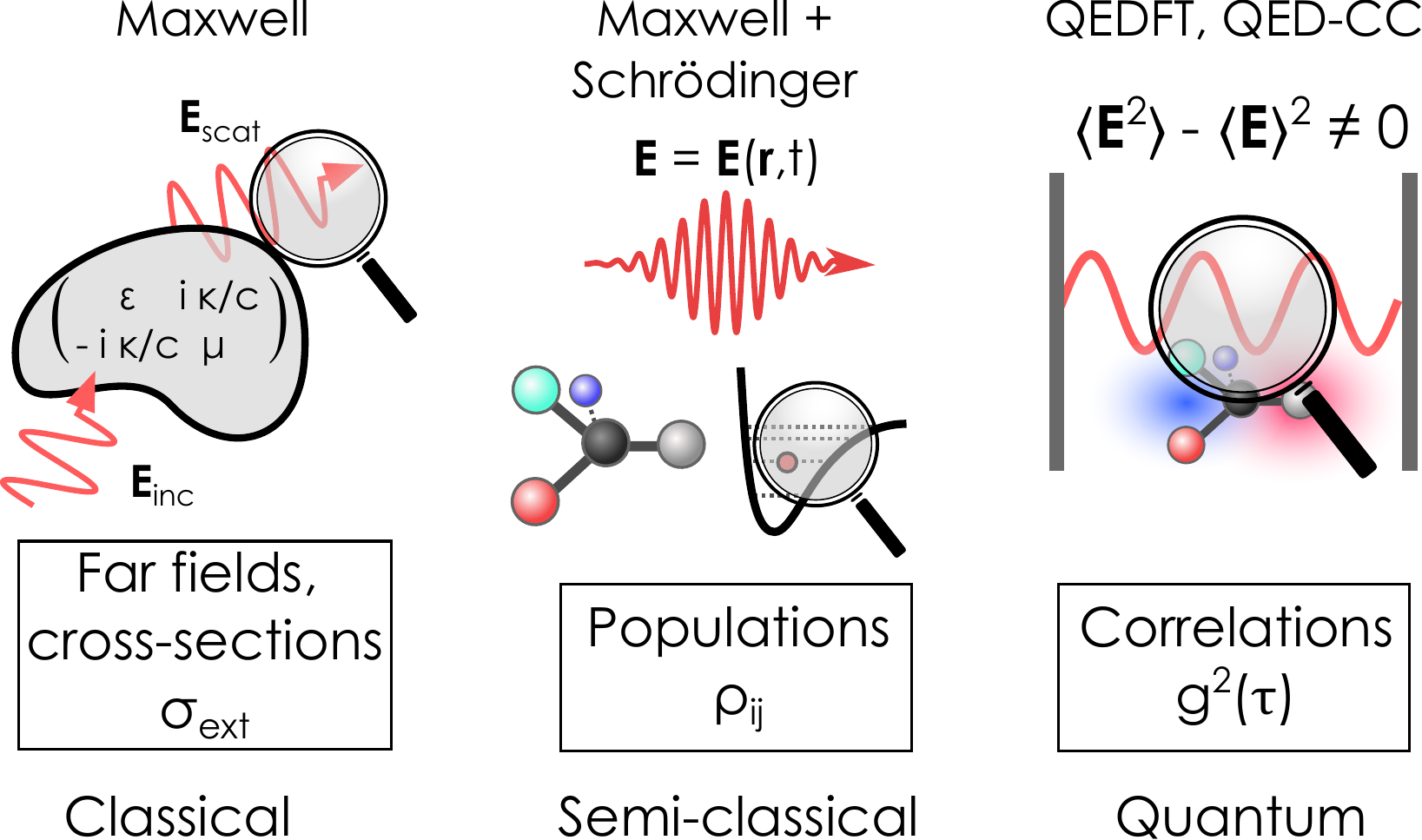}
\caption{\textbf{Summary of classical and quantum theoretical methods for description of strong chiral light-matter interaction.} Left: description/modelling of chiral systems using Maxwell's equations in terms of macroscopic constitutive parameters gives access to spectroscopic far-field quantities, such as cross-sections. Middle: semi-classical description using combined treatment with Maxwell's equations for the field and Schr\"odinger equation for matter gives access to, e.g., population dynamics of the matter encoded by its density matrix with feedback on the optical fields. Right: full quantum description of chiral light-matter interaction gives access to quantum correlated features, such as ground-state changes and combined statistics.
}
\label{fig8}
\end{figure*}

Simplified analytical models are mandatory to build intuition -- the foundation for innovative developments and starting point for sophisticated descriptions. However, a faithful description of experimental reality, which provides in the long run the necessary microscopic understanding, will inevitably require sophisticated microscopic theory. How complex such a microscopic description has to be will depend heavily on the system at hand and the relevant observable, as sketched in Fig.~\ref{fig8}. %We can separate the discussion conceptually into two parts, one in which the light-matter interaction is fully self-consistent but classical, and one in which the chiral electromagnetic fields are quantized.
%\paragraph{Quantum matter and classical light}
The value of the classical approximation for light should not be underestimated. Providing Maxwell's equations with an adequate polarizability will often suffice to obtain accurate far-field spectra.
As an example, consider the typical Hopfield system with a set of quantum harmonic oscillators describing molecular dipoles and a quantized photonic mode. In the linear regime, its spectrum is identical to the purely classical solution obtain from Maxwell's equation \cite{Todorov2012,DeLiberato2017}.
%\Cadd{[@Denis: Could you please add here a figure where you plot the Hopfield and Maxwell solution (both identical) ]}
Clearly, Maxwell's equation does not provide access to the microscopic dynamic governed by quantum mechanics -- the moment we intend to control the material we are bound to consider its full complexity. However, it is often sufficient to combine quantum matter with classical light. 

%\Cadd{Add some simple explanatory equations or a simple figure or so to make the differences clear (full Maxwell, quantum matter+microscopic maxwell, quantum optical models, full quantum matter + quantum light}

The complexity of quantum matter imposes a necessity for efficient descriptions such as time-dependent density-functional theory\cite{gross1984} which can be combined with a self-consistent propagation of classical light \cite{schafer2021shortcut,schafer2022emb,tancogne2019octopus,noda2019salmon}. This approach remains accurate in predicting the eigenstates of the combined system even in the case of strong interaction with optically confined modes\cite{flick2018light}. %It should be noted again that the dipolar moments are defined with respect to the center of mass of each molecule. This demands careful consideration when many molecules are coupled collectively and approaches, such as the embedding radiation-reaction (ERR) ansatz \cite{schafer2022emb}, which separate the combined system into subsystem, might be beneficial.
Naturally, the major limitation of this approach is that quantum fluctuations of the electromagnetic field are not considered. While mixed quantum-classical approaches are partially able to compensate for the statistical nature of quantum fluctuations \cite{hoffmann2019capturing,chen2019ehrenfest,hoffmann2019benchmarking,doi:10.1021/acs.jpclett.1c00158}, much remains to be desired. Changes in the correlated ground-state are for instance entirely caused by quantum fluctuations. % and demand approaches that will be highlighted in the following.

%\paragraph{Quantum matter and quantum light}
Following the canonical quantization procedure \cite{Craig1999,andrews2018perspective}, the photonic fields are represented as a sum over many quantum harmonic oscillators. This vastly enlarges the already excessive size of the material Hilbert space and renders a direct solution of the associated Schr\"odinger equation impossible for anything remotely realistic. Typical coping strategies assume, for instance, a strong simplification of the material.\cite{dicke1954,kockum2019ultrastrong,schafer2022chiral} In general, however, this approach is best suited to collect intuition and extensions to a more faithful representation of the material become important when working with e.g. chemical reactions. The complexity of light and matter is dominated by the complexity of the material. For weak light-matter interaction, the associated Fockspace can be efficiently truncated. Ultra-strong interactions require effective photon-free and non-perturbative approaches\cite{schafer2021making,ashida2021}.

If the complexity of the material represents the major limitation, the solution should be to adjust existing many-body techniques to include quantized light-matter coupling. Two such approaches are quantum-electrodynamical density-functional theory (QEDFT)\cite{ruggenthaler2014quantum,tokatly2013,flick2017atoms,schafer2021making} and QED coupled-cluster (QED-CC)\cite{haugland2020coupled,mordovina2020polaritonic,fregoni2021strong,deprince2021cavity}. Their QED adapted form inherits the fundamental strength and weaknesses of their respective quantum mechanical counterparts. While this novel field of \textit{ab initio} QED\cite{schafer2018ab} is still in its infancy, already the initial development provides access to real-space resolved Lamb-shifts\cite{flick2018ab}, modification of chemical reactivity \cite{schafer2021shining}, and intermolecular interactions \cite{haugland2021intermolecular}. Chiral polaritonics calls now for an intensified development in two directions.
%It should be noted that none of the above techniques is well suited to tackle all challenges in the field and a resourceful scientific approach should select its tools on the basis of requirement and cost. A reliable and unprejudiced \textit{ab intio} framework is, without question, an integral part of the envisioned tool-set.

%The major limitation of the aforementioned techniques is the time-consuming development process and high computational cost. Every development step has to conserve the effective formulation of the chosen technique, ensure stability and gauge-invariance of the non-linear algorithm, and keep the computational cost in check. The necessary attention to each detail requires often to scrutinize commonly employed approximations\cite{schaffer2020relevance} and reinterpret existing problems\cite{schafer2021making}. 

\paragraph{Beyond the electric dipole:}

While many optical effects, such as Mie-scattering, imply matter to be of the size of the optical wavelength, the molecular constituents of the material are of much smaller size and the electric dipole commonly dominates the interaction.
This convenient feature led to a clear dominance of the electric dipole approximation for the interaction between quantized fields with matter -- to a point that the implications of this approximation had been partially forgotten\cite{Schafer2020relevance,stokes2022implications}. Now, within the realm of chiral strong-coupling, it will be required to remedy this development. It seems natural to extend the description simply to the next order in the multipolar expansion\cite{Craig1999} and the multipolar framework is known to perform well in the single mode limit of strong coupling\cite{stokes2022implications}.
In addition, full minimal coupling descriptions could complement this process, providing an efficient sanity check via the restrictions enforced by gauge invariance. The latter is only intact for the full Hilbert space. Furthermore, a resolution of the photonic momentum might question the single-mode approximation, bringing the field in closer contact to Casimir physics. Riso et al. have recently taken first steps in this direction by minimally coupling a single cavity mode to chiral molecules described by coupled-cluster theory\cite{riso2022strong}. An alternative direction is based on extensions of (current-)density functional theory\cite{ruggenthaler2014quantum,schafer2021making,jestadt2019light}.

\paragraph{Scaling up:}

The most elegant or rigorous theory is useless if unable to predict and understand the problem at hand. Thus, we are facing the question: Given a satisfying theoretical description of a chiral molecule, how do we account for the chiral environment comprising chiral cavities and nanoplasmonic particles?
The recent debate around the theoretical description of ground-state polaritonic chemistry\cite{garcia2021manipulating,simpkins2021mode,sidler2022perspective,Mandal2022review} illustrates the limitation of oversimplified models. While \textit{ab initio} QEDFT calculations\cite{schafer2021shining} have shown qualitative agreement with experiments, much remains to be understood and the simplified description of the photonic system to be scrutinized. Collective light-matter interaction and complex environments represent now the next challenge\cite{schafer2022emb}.
Conceptually, this is closely related to other multi-scale problems, such as impurities in condensed matter\cite{RevModPhys.68.13} or active centers in biological structures\cite{neugebauer2009subsystem,levitt2014birth}. 
While parallelization strategies can account, under severe approximations, for a few hundred of molecules\cite{luk2017multiscale}, mesoscopic ensembles and realistic macroscopic environments will only become available when efficient embedding strategies are employed\cite{schafer2022emb,PhysRevLett.125.233603,spano2020exciton,cui2022collective,mctague2022non,doi:10.1063/5.0095552,sanchez2021few}. Lastly, plasmonic structures will take a prominent role for chirality\cite{kotov2021chiral}, especially due to their capability to enhance the inherently weak chirality (see Sec.~\ref{sec:perspective}). Charge migration and the microscopic surface structure renders simple eigenmode quantization strategies questionable if our chiral molecule of interest is in close proximity to the nanoparticle. Density-functional theory presents a reliable choice where first connections to optical cavities are already available\cite{schafer2021shortcut}. The explicit simulation of sizeable plasmonic structure is, however, too costly. Mixed descriptions along the lines of quantum hydrodynamics\cite{doi:10.1063/1.5003910} or the embedding of quantum-dynamics into classical permittivity for larger fractions of the system\cite{schafer2022emb,doi:10.1021/acs.jpcc.2c02209} could help to alleviate some of the computational weight necessary to describe realistic plasmonic systems.

%%%%%%%%%%%
\section{Conclusions}
To conclude, we have presented our vision of a potentially new area of nanophotonics - chiral polaritonics.
Guided by the notion of chirality of \elm field and matter, we established the general requirements to resonant optical systems in which chiral polaritonic eigenstates may emerge.
Recent theoretical efforts discussed in this Perspective already indicate that chiral polaritonic systems may feature non-trivial optical phenomena, where the interplay of light and matter chirality is of key importance for determining the eigenfunctions of the system, as well as its response to external \elm fields.

We have speculated on the potential of a collection of novel effects that can be enabled by strong coupling between chiral light and matter. 
Established directions for the utilization of chirality %, in our opinion, 
involve plasmonic or biological nanostructures\cite{kotov2021chiral}, the here proposed usage of chiral polaritonics adds further versatility. The addition of 'light', being the fundamental reason for dispersive/van-der-Waals interactions, is more than a mere hurdle -- it provides a new opportunity for the description and control of chirality.
Finally, development of existing and new theoretical tools might be necessary in order to accurately describe various observables of a strongly coupled chiral system, such as far-field response functions, molecular populations, and quantum correlations.
Due to the multifaceted nature of the problem, a coordinated effort of researchers with backgrounds in classical electromagnetism, quantum optics, and material science will be needed in order to succeed in the hunt for chiral polaritons.

%\Cadd{[The following 2 sentences might be better placed in the conclusion.]}

\begin{acknowledgement}
%Authors acknowledge fruitful discussion with ...
The work was supported by the Russian Science Foundation (21-72-00051). The work of M.V.G. was supported by the Ministry of Science and Higher Education of the Russian Federation within the State assignment of FSRC ``Crystallography and Photonics'' RAS. C.S. acknowledges funding by the Swedish Research Council (VR) through Grant No. 2016-06059. D.G.B. acknowledges support from BASIS Foundation (grant 22-1-3-2-1).
\end{acknowledgement}

\bibliography{chir_polaritons}

\end{document}